\documentclass[iop]{emulateapj-rtx4}

\begin{document}

\title{A new infrared color criterion for the selection of $0<z<7$ AGN: application to deep fields and implications for JWST surveys.}

\author{H. Messias\altaffilmark{1}\altaffilmark{2}, J. Afonso\altaffilmark{1}, M. Salvato\altaffilmark{3}\altaffilmark{4}, B. Mobasher\altaffilmark{5}, A. M. Hopkins\altaffilmark{6}}
\email{hmessias@oal.ul.pt}

\altaffiltext{1}{Centro de Astronomia e Astrof\'{i}sica da Universidade de Lisboa, Observat\'orio Astron\'omico de Lisboa, Tapada da Ajuda, 1349-018 Lisboa, Portugal.}
\altaffiltext{2}{Departamento de Astronom\'ia, Av. Esteban Iturra 6to piso, Facultad de Ciencias F\'isicas y Matem\'aticas, Universidad de Concepci\'on, Chile.}
\altaffiltext{3}{Max-Planck for Extraterrestrial Physics , Giessenbachstrasse 1, Garching, 85748 Germany.}
\altaffiltext{4}{Excellence Cluster, Boltzmann Strasse 2,  Garching, 85748 Germany.}
\altaffiltext{5}{University of California, 900 University Ave., Riverside, CA 92521, USA.}
\altaffiltext{6}{Australian Astronomical Observatory, P.O. Box 296, Epping, NSW 1710, Australia.}

\begin{abstract}
It is widely accepted that observations at mid-infrared (mid-IR) wavelengths enable the selection of galaxies with nuclear activity, which may not be revealed even in the deepest X-ray surveys. Many mid-IR color-color criteria have been explored to accomplish this goal and tested thoroughly in the literature. Besides missing many low-luminosity active galactic nuclei (AGN), one of the main conclusions is that, with increasing redshift, the contamination by non-active galaxies becomes significant (especially at $z\gtrsim2.5$). This is problematic for the study of the AGN phenomenon in the early Universe, the main goal of many of the current and future deep extra-galactic surveys. In this work new near- and mid-IR color diagnostics are explored, aiming for improved efficiency --- better completeness and less contamination --- in selecting AGN out to very high redshifts. We restrict our study to the \textit{James Webb Space Telescope} wavelength range (0.6--27\,$\mu$m). The criteria are created based on the predictions by state-of-the-art galaxy and AGN templates covering a wide variety of galaxy properties, and tested against control samples with deep multi-wavelength coverage (ranging from the X-rays to radio frequencies). We show that the colors $K_s-[4.5]$, $[4.5]-[8.0]$, and $[8.0]-[24]$ are ideal as AGN/non-AGN diagnostics at, respectively, $z\lesssim1$, $1\lesssim{z}\lesssim2.5$, and $z\gtrsim2.5-3$. However, when the source redshift is unknown, these colors should be combined. We thus develop an improved IR criterion (using $K$ and IRAC bands, KI) as a new alternative at $z\lesssim2.5$. KI does not show improved completeness (50--60\% overall) in comparison to commonly used IRAC-based AGN criteria, but is less affected by non-AGN contamination (revealing a $>$50--90\% level of successful AGN selection). We also propose KIM (using $K$, IRAC, and MIPS-24\,$\mu$m bands, KIM), which aims to select AGN hosts from local distances to as far back as the end of reionization ($0<z\lesssim7$) with reduced non-AGN contamination. However, the necessary testing-constraints and the small control-sample sizes prevent the confirmation of its improved efficiency at $z\gtrsim2.5$. Overall, KIM shows a $\sim$30--40\% completeness and a $>$70--90\% level of successful AGN selection. KI and KIM are built to be reliable against a $\sim$10--20\% error in flux, are based on existing filters, and are suitable for immediate use.
\end{abstract}

\keywords{galaxies: active; galaxies: high-redshift; Infrared: galaxies}

\section{Introduction}

The active galactic nucleus (AGN) phenomenon in galaxies is currently the focus of a substantial amount of attention, due to its role in altering the path of galaxy evolution \citep[e.g.,][]{Granato04,Springel05,Croton06,Hopkins06,Bower06,Somerville08}. The quest to find these sources and the techniques used in identifying them are constantly shaping our understanding of the underlying physics of AGN, which are currently believed to be the result of mass accretion onto super massive black holes at the very centre of galaxies \citep[e.g.,][]{Laor99,Melia01}. To identify AGN, a miriad of techniques have been developed, spanning essentially the full electromagnetic spectrum (e.g., X-rays, \citealt{Szokoly04}; optical, \citealt{Richards02}; infrared, see references below; radio, \citealt{Tielens79,Chambers96}).

A wide variety of both intrinsic and observational effects, make the identification of large, statistically robust, but also complete and reliable, populations of AGN galaxies challenging. Such effects encompass obscuration, accretion rate, variability, host galaxy light, viewing angle, and more. Different aspects of AGN radiation may be observed across the entire electromagnetic spectrum \citep[from direct light at high energies to synchrotron radiation dominating at radio frequencies,][]{TreisterUrry11}. It is common, though, to observe and analyze aspects of AGN activity limited to only one spectral regime. This has been verified in extreme radio sources \citep{Afonso06,Simpson06,Morganti11,Norris11} and in some X-ray sources \citep[e.g.,][]{Loewenstein01,Ho03,Maiolino03,Polletta06}. These specific sources, show an ultraviolet-to-infrared spectrum with no evidence for AGN activity, and some of these AGN may not even present any counterpart at those wavelengths (as a result of, for instance, extreme obscuration). Even more generally, there is strong evidence for a large population of obscured AGN, which still remain unidentified at high-energy spectral-bands \citep[e.g.,][]{Comastri01,Ueda03,Gilli04,Worsley04,Worsley05,TreisterUrry05,Martinez05,Draper09}. If large dust column densities are the driver for such extreme obscuration (instead of just gas), then infrared (IR) wavelengths, at which the obscuring dust emits, are an obvious choice in the search for such extreme objects.

Based on data from the Infrared Astronomical Satellite \citep[IRAS,][]{Neugebauer84}, \citet{deGrijp85} and \citet{Miley85} realized that simple spectral-index or color-color diagnostics could be efficient ways to select Seyfert- and QSO-type galaxies. More recently, with the launch of \textit{Spitzer} \citep{Werner04}, the same techniques were applied at the shorter wavelengths ($3.6-8\,\mu$m) of the Infrared Array Camera \citep[IRAC,][]{Fazio04}, where hotter dust temperatures are probed. \citet{Lacy04}, \citet{Hatziminaoglou05}, and \citet{Stern05} proposed different IRAC color-color plots to select AGN \citep[see also][]{Sajina05}, while \citet{Alonso06} and \citet{Donley07} proposed spectral-index diagnostics to select sources revealing power-law ($f\propto\nu^\alpha$) spectral energy distributions (SEDs) characteristic of AGN \citep{Neugebauer79,Elvis94,Ivezic02}. Later, by comparison of the optical spectral regime with IR wave-bands, other techniques were proposed. Either through SED fitting \citep{Daddi07} or extreme $R-[24]$ color cuts at different 24\,$\mu$m fluxes \citep{Polletta08,Dey08,Fiore08,Donley10}, the selected galaxies are too red and bright for a normal star-forming SED, and hence indicative of AGN activity. 

However, some critical shortcomings have been highlighted specially regarding the IRAC color-color AGN selection technique \citep[e.g.,][]{Barmby06,Donley08,Eckart10,Donley12}. Having been developed based on the shallow first generation IRAC surveys, these techniques fail in deeper IRAC surveys, where higher redshift non-AGN sources pollute the adopted selection criteria \citep{Barmby06,Papovich06,Donley08,Messias10,Donley12}. Also, this technique is strongly biased against selecting low-luminosity AGN \citep{Treister06,Cardamone08,Donley08,Eckart10,Petric11,Donley12}, an evolutionary phase where AGN are believed to spend most of their life span \citep[e.g.,][]{Hopkins05,Hopkins06,Fu10}. Furthermore, while the IR selection techniques return a variety of AGN populations, there seems to be a bias toward the selection of unobscured over obscured sources \citep{Stern05,Donley07,Cardamone08,Eckart10}, which, strikingly, is the opposite of the goal that drives IR AGN selection in the first place.

More recently, alternative diagnostics have been proposed either tuned to specific samples \citep[e.g.,][]{Garn09,Messias10,Donley12}, or to newly available filter sets \citep{Assef10}, or in combination with other techniques \citep{Richards09,Edelson12}. Such work is crucial and the quest for revised and/or improved color-color criteria should continue, as it is important to identify key filter sets in order to maximize survey efficiency, as well as current \citep[e.g.,][]{Salvato09,Salvato11,Hainline11,Pozzi12} and future SED fitting procedures aiming to break down the strong degeneracy between AGN and non-AGN SEDs.

In this work, we explore new IR color selections with the goal to achieve improved efficiency (better completeness and less non-AGN contamination). In Section~\ref{sec:samp} different possibilities for the mechanisms behind the IR emission are discussed, and the new criteria are presented in Section~\ref{sec:newcrit}. A series of tests of these new criteria are explored in Section~\ref{sec:ctrlsp} using a broad set of deep multi-wavelength control samples (selected at wavelengths ranging from X-ray to radio). Implications for and expectations from \textit{James Webb Space Telescope} (\textit{JWST}) surveys are highlighted in Section~\ref{sec:impjwst}, followed by the conclusions of this work in Section~\ref{sec:conc}.

Throughout this paper we use the AB magnitude system\footnote{When necessary the following relations are used:
(K, [3.6], [4.5], [5.8], [8.0])$_{AB}$ = (K, [3.6], [4.5], [5.8], [8.0])$_{Vega}$ + (1.841, 2.79, 3.26, 3.73, 4.40) \citep[][and http://spider.ipac.caltech.edu/staff/gillian/cal.html]{Roche03}.}, assuming a $\Lambda$CDM model with H$_{0} = 70$ km s$^{-1}$ Mpc$^{-1}$, $\Omega_{M} = 0.3$, $\Omega_{\Lambda} = 0.7$.

\section{Distinguishing AGN from non-AGN IR contributions} \label{sec:samp}

\begin{figure}
\plotone{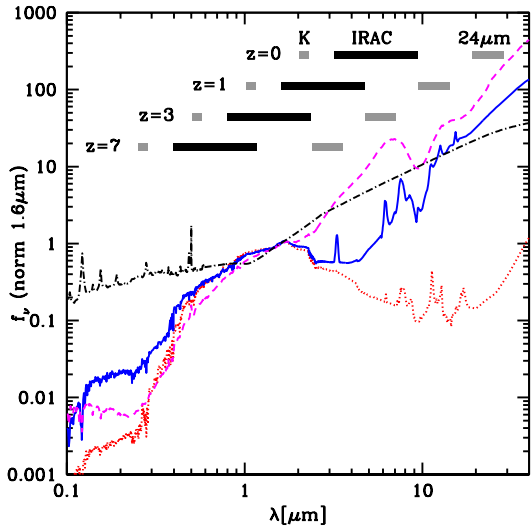}
\caption{Examples of galaxy templates taken from the SWIRE Template Library \citep{Polletta07} and flux normalized at 1.6$\mu$m: S0 (early type galaxy, red dotted line), M82 (starburst galaxy, blue solid line), IRAS 19254-7245 (a hybrid source, magenta dashed line), and a type-1 QSO (AGN, black dot-dashed line). The shaded regions show what rest-frame wavelength the K, IRAC, and MIPS 24${\mu}m$ filters will be observing depending on the redshift.\label{fig:sed}}
\end{figure}

The SEDs of stellar and/or star-formation (SF) dominated systems (henceforth referred to as normal galaxy SEDs) have some distinctive characteristics, allowing the separation of this population from AGN host galaxies through the use of IR colors alone. Figure~\ref{fig:sed} illustrates a few examples using galaxy templates taken from the SWIRE Template Library \citep{Polletta07}. In normal galaxy SEDs, a dip in the 1--6\,$\mu$m rest-frame range is expected. The overall blackbody emission from the stellar population combined with the minimum in the opacity of the H$^{-}$ ion in stellar photospheres, peaks at $\sim1.6\,\mu$m, which is generally observed, as is the CO absorption at 2.35--2.5\,${\mu}$m from red supergiants. Emission from hot dust surrounding stars in the asymptotic giant branch (AGB) phase is expected in young systems ($\lesssim1\,$Gyr) and to peak at $2-5\,\mu$m \citep{Sajina05,Maraston05,Henriques11}. Furthermore, the strength of the polycyclic aromatic hydrocarbons (PAH) features, seen mostly beyond 6\,$\mu$m, increases with star formation activity and metallicity \citep{Calzetti07,Engelbracht08,Hunt10}. It is in this spectral region (1--6\,${\mu}$m) that the difference between normal galaxies and AGN dominated SEDs is the greatest. The existence of an AGN is frequently accompanied by a power-law continuum \citep[$f_{\nu}~\propto~{\nu}^{\alpha}$, e.g.,][]{Neugebauer79,Elvis94,Ivezic02} rising in flux beyond $\sim1\,\mu$m. This is a consequence of X-ray-to-optical AGN radiation being reprocessed to IR wavelengths by dust surrounding the central region of an active galaxy \citep{Sanders89,Sanders99,PierKrolik92}. This feature, being different from the dip observed in normal galaxy SEDs, is unique for AGN hosts and enables their selection through SED inspection.

\subsection{The template set} \label{sec:templates}

The templates used throughout this paper come from published work as follows: 10 templates covering early to late galaxy types, five starbursts, four hybrids\footnote{By \textit{hybrids} we refer to SEDs simultaneously including stellar/SF and AGN emission.}, and six AGN, all from \citet{Polletta07}; nine starburst Ultra-Luminous IR Galaxies (ULIRGs) from \citet{Rieke09}; one blue starburst and 18 hybrid SEDs from \citet{Salvato09}; and one extremely obscured hybrid from \citet{Afonso01}.

All templates have been fully characterized in the papers which introduce them (see above). We assume that they sample adequately the color-$z$ space, based on the results by \citet{Salvato09,Salvato11}. In these works, the templates above (except those from \citealt{Afonso01}, and \citealt{Rieke09}) were used to compute reliable photometric redshift estimates (reaching an accuracy of $\sigma_{\Delta z/(1+z_{\rm spec})}\sim0.015$, where $\Delta z=|z_{\rm spec}-z_{\rm phot}|$) for both bright, nearby objects and faint, high-$z$ sources ($z<5.5$). By further adopting the models from \citet{Afonso01} and \citet{Rieke09}, the whole template set aims to account for a wide range of mixtures between star-formation and AGN emission in galaxy SEDs and their evolution across cosmic time, as evidenced by observations \citep[e.g.,][]{Farrah07,Papovich07,Rigby08,Pope08,Fiore08,Donley10,Fu10,Elbaz11}. This is a valid assumption given that local templates are successful in fitting some of the most extreme high redshift sources (for instance, the case of Arp220 as a local analog of HR10, an extremely red galaxy at $z=1.44$, \citealt{HuRidgway94} and \citealt{Elbaz02}, or M82 as an analog for star-formation dominated sub-millimeter galaxies, \citealt{Pope08}).

Five ULIRGs (IRAS sources 12112+0305, 14348-1447, 17208-0018, 20551-4250, and 22491-1808) which were previously considered as pure-starbursts in the literature, are labelled as hybrid sources in this work. This is based on the findings by \citet{Veilleux09} and the review by \citet{SaniNardini11}, where IR spectroscopy is shown to reveal significant AGN contribution (6--30\% at 8--1000\,$\mu$m, 8--40\% at 5--8\,$\mu$m) in these ULIRGs.

Table~\ref{tab:temp} lists the adopted template set. The SED templates are organized in four groups: (a) Early to Late-type galaxies, (b) Starbursts, (c) Hybrids, and (d) AGN. The following investigation will focus on how these groups populate near-to-mid IR color-color spaces, aiming to separate the AGN/Hybrid population, (c) and (d) above, from normal galaxies, i.e., (a) and (b).

\begin{flushleft}
\begin{deluxetable}{ccc}
\tabletypesize{\scriptsize}
\tablecaption{List of templates.\label{tab:temp}}
\tablewidth{0pt}
\tablehead{
\colhead{Class} & \colhead{template} & \colhead{Ref\tablenotemark{a}}
}
\startdata
Early/Late & Ell13 & [2] \\
& Ell2 & '' \\
& Ell5 & '' \\
& S0 & '' \\
& Sa & '' \\
& Sb & '' \\
& Sc & '' \\
& Sd & '' \\
& Sdm & '' \\
& Spi4 & '' \\
\hline
Starburst & M82 & [2] \\
& Arp 220 & '' \\
& NGC 6090 & '' \\
& ESO 0320-g030 & [4] \\
& NGC 1614 & '' \\
& NGC 2369 & '' \\
& NGC 3256 & '' \\
& NGC 4194 & '' \\
& Zw049.057 & '' \\
& CB1\_0\_LOIII4 & [3] \\
\hline
Hybrid & IRAS 12112+0305 & [4,5] \\
& IRAS 14348-1447 & '' \\
& IRAS 17208-0018 & ''\tablenotemark{b} \\
& IRAS 22491-1808 & [2,5] \\
& IRAS 20551-4250 & [2,6] \\
& IRAS 19254-7245 & '' \\
& NGC 6240 & '' \\
& Seyfert 2.0 & '' \\
& Seyfert 1.8 & '' \\
& ERO & [1] \\
& pl\_I22491\_10\_TQSO1\_90 & [3] \\
& pl\_I22491\_20\_TQSO1\_80 & '' \\
& pl\_I22491\_30\_TQSO1\_70 & '' \\
& I22491\_40\_TQSO1\_60 & '' \\
& I22491\_50\_TQSO1\_50 & '' \\
& I22491\_60\_TQSO1\_40 & '' \\
& I22491\_70\_TQSO1\_30 & '' \\
& I22491\_80\_TQSO1\_20 & '' \\
& I22491\_90\_TQSO1\_10 & '' \\
& S0\_10\_QSO2\_90 & '' \\
& S0\_20\_QSO2\_80 & '' \\
& S0\_30\_QSO2\_70 & '' \\
& S0\_40\_QSO2\_60 & '' \\
& S0\_50\_QSO2\_50 & '' \\
& S0\_60\_QSO2\_40 & '' \\
& S0\_70\_QSO2\_30 & '' \\
& S0\_80\_QSO2\_20 & '' \\
& S0\_90\_QSO2\_10 & '' \\
\hline
AGN & Mrk 231 & [2] \\
& BQSO1 & '' \\
& QSO1 & '' \\
& TQSO1 & '' \\
& QSO2 & '' \\
& Torus & '' \\
\enddata
\tablenotetext{a}{Reference work from which the template was retrieved, and, if present, reference work revealing the presence of hidden or non-dominant AGN (right column): [1] \citet{Afonso01}, [2] \citet{Polletta07}, [3] \citet{Salvato09}, [4] \citet{Rieke09}, [5] \citet{Veilleux09}, [6] \citet{SaniNardini11}.}
\tablenotetext{b}{\citet{Veilleux09} do not present results for IRAS 17208-0018 specifically, but \citet[][the work where that template comes from]{Rieke09} use Spitzer-IRS spectroscopy on IRAS 22491-1808 \citep[studied by][]{Veilleux09} to complete ISO observations on IRAS 17208-0018, invoking similarities between the two in the 5--11\,$\mu$m range, where the analysis by \citet{Veilleux09} is done.}
\end{deluxetable}
\end{flushleft}

\vspace{1cm}

\section{The new approach} \label{sec:newcrit}

\subsection{An enhanced wedge diagram: the KI criterion}\label{sec:ki}

\begin{figure*}
\plottwo{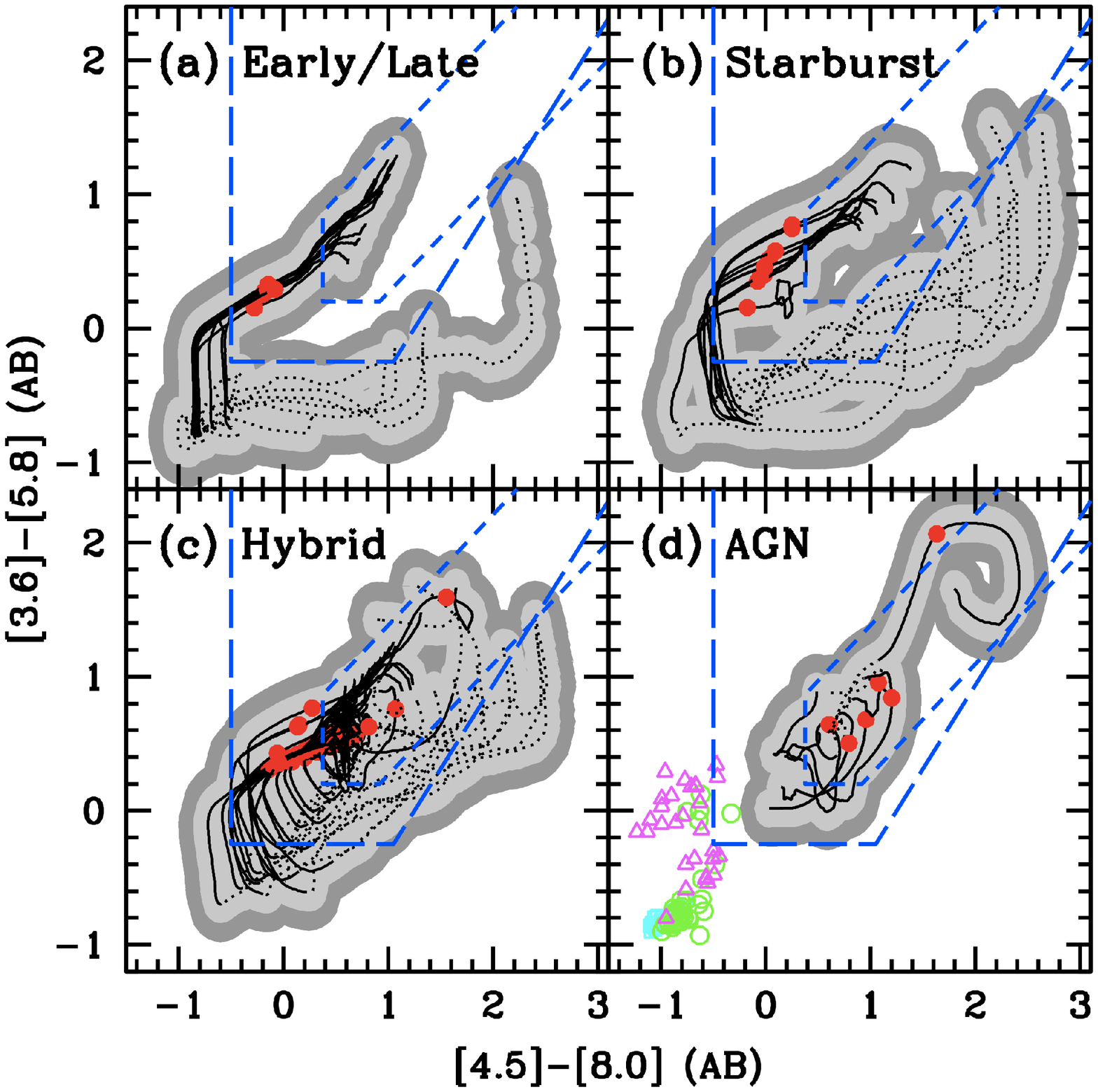}{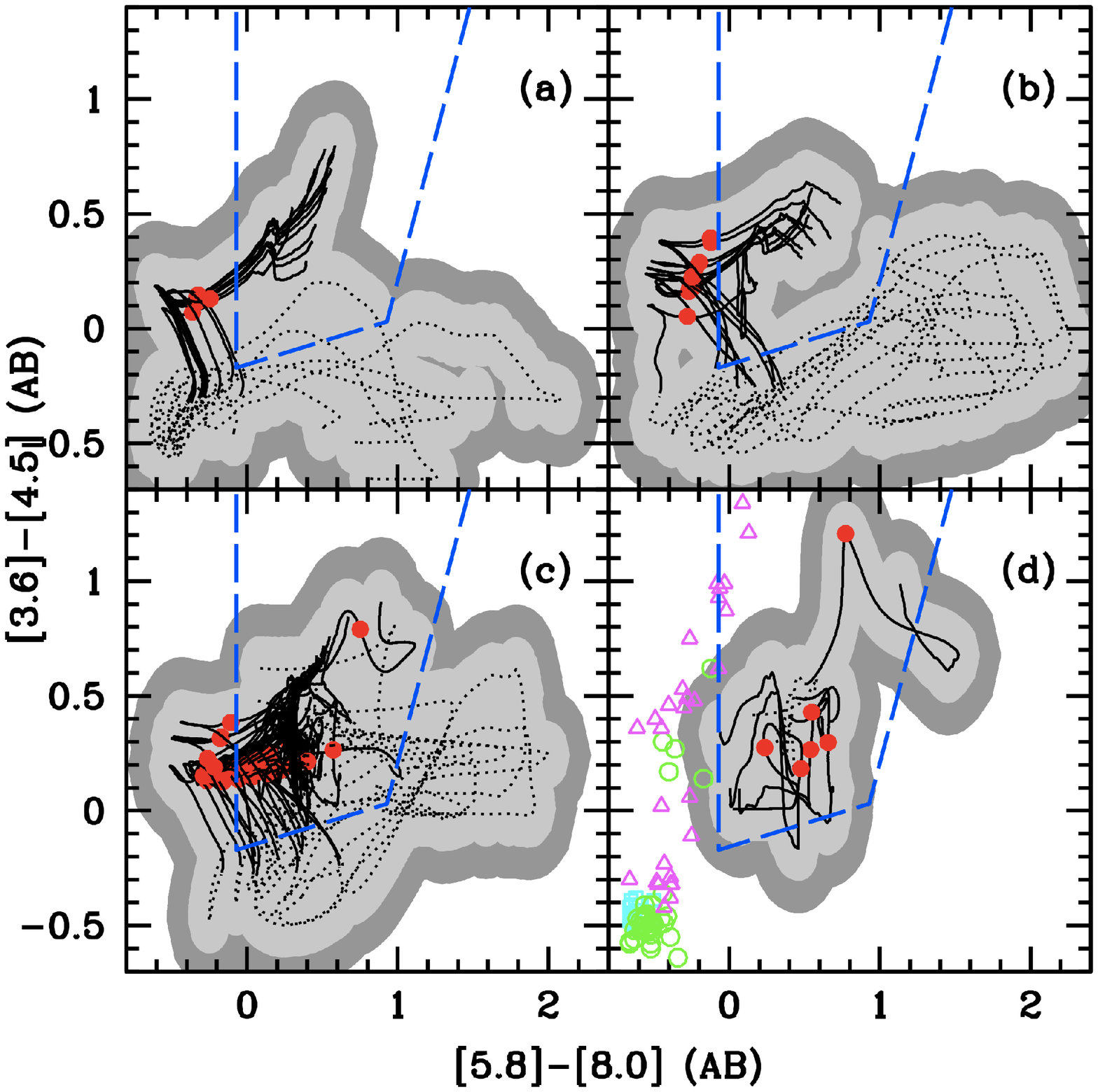}
\caption{Model color tracks displayed in the L07 (left plot) and S05 (right plot) criterion color-color space. Long-dashed blue lines correspond to the boundaries proposed in those works for the selection of AGN, while short-dashed blue lines in the left plot correspond to the color boundaries proposed by \citet{Donley12}. Each panel presents a specific group: (a) Early/Late, (b) starburst, (c) Hybrid and (d) AGN. The dotted portion of the tracks refers to the $0<z<1$ redshift range, and the solid to $1\leq{z}\leq7$. Red circles along the lines mark $z=2.5$. The light and dark-grey regions show the photometric scatter due to a magnitude error of, respectively, 0.1 and 0.2 in the considered bands (equivalent to $\sim10\%$ and $\sim20\%$ error in flux, respectively). Dwarf stars \citep{Patten06} are shown for reference: M-dwarfs appear as open cyan squares, L-dwarfs as open green circles, and T-dwarfs as open magenta triangles to show where these red point-like cool stars appear.\label{fig:c1324}}
\end{figure*}

In Figure~\ref{fig:c1324} the color tracks (spanning the range $0<z<7$) for the template SEDs considered are presented on the L07 (left plot) and S05 (right plot) criteria color-color spaces. The criterion recently proposed by \citet[][D12 henceforward]{Donley12} is also shown in the L07 color-color space, with the caveat that the D12 criterion requires a monotonically increasing red SED in addition to the color cuts shown in the figure. These criteria consider the 3.6--8.0\,$\mu$m observed-frame (the IRAC-frame) to select AGN candidates. This is the spectral range where AGN-dominated SEDs are better distinguishable from normal galaxy SEDs at rest-frame wavelengths, meaning that redshift will play a role in the selection efficiency. In both color-color spaces, the nominal AGN regions encompass most of the AGN and hybrid tracks for a large range of redshifts, as they were built to do (the D12 criterion focus only on the most SED-dominant AGN).

We note, however, that the use of a very diverse SED template set already shows some shortcomings of these diagnostic plots. In both plots, the two upper panels (early/late and starburst galaxies) show a significant contamination of the nominal AGN region of L07 and S05 (and D12 to a smaller degree) by normal (non-AGN) galaxies. Each criterion is differently affected at different redshifts. Even when ignoring photometric scatter (shaded regions in the figures), one observes non-AGN contamination in at least one criterion throughout the whole $0<z<7$ range. This has been noted by previous studies testing the L07, S05, and D12 criteria \citep{Barmby06,Donley08,Cardamone08,Donley12}, but it is also found for other IR techniques at higher redshifts \citep[$z\gtrsim2$; e.g.,][]{Pope08,Donley08,Donley10,Fadda10,Narayanan10}. While at low-redshifts ($z\lesssim2-3$), the contamination is due to strong PAH emission as well as extreme obscuration differently affecting each IRAC channel, at higher redhifts ($z\gtrsim2-3$), the contamination is mostly driven by the limitation in observed wavelengths (the IRAC frame, $3.6<\lambda[\mu{\rm m}]<8$). At these redshifts, IRAC bands mostly probe the rest-frame $\lambda\lesssim1.6\,\mu$m, where obscured star-forming systems (frequently identified at those redshifts) present red colors mimicking those of AGN. The clear implication is that any AGN criteria restricted to the IRAC-frame will never be able to disentangle AGN from non-AGN populations at $z\gtrsim3$. In Figure~\ref{fig:c1324}, the highest redshift where non-AGN templates are restricted to lying outside the S05 and D12 selecting regions is $z\sim2.5$. At higher redshifts, L07, S05, and D12 are all contaminated to the same degree. At $z<2.5$, the D12 criteria is the least contaminated \citep[as shown in][]{Donley12}, but being restricted to such a narrow color-color space, it will miss a significant number of low- and intermediate-luminosity AGN ($\rm{\log(L_X[erg\,s^{-1}])<44}$). In addition, cool dwarf stars may fall inside or close to the L07 and S05 regions, thus being potential (point-like) contaminants.

In order to enhance these wedge diagrams, one can extend the wavelength coverage to different wavebands. This is obviously outside the IRAC framework behind the original definition of such wedge diagrams, but suits the larger \textit{JWST} wavelength coverage. For instance, by considering shorter wavelengths ($<3\,\mu$m), one is probing a spectral region mostly dominated by stellar emission\footnote{However, AGN emission may also dominate at such short IR wavelengths. See, for example, \citet{Assef11}, \citet{Hainline11}.} (see Figure~\ref{fig:sed}). Such a scenario is an advantage as we now compare a stellar dominated wave-band with one at longer wavelengths (e.g., the IRAC-frame) that has contribution either from stellar or AGN light. Such a comparison will yield a large color dispersion, which is ideal for the separation of the two types of systems.

\begin{figure}
\plotone{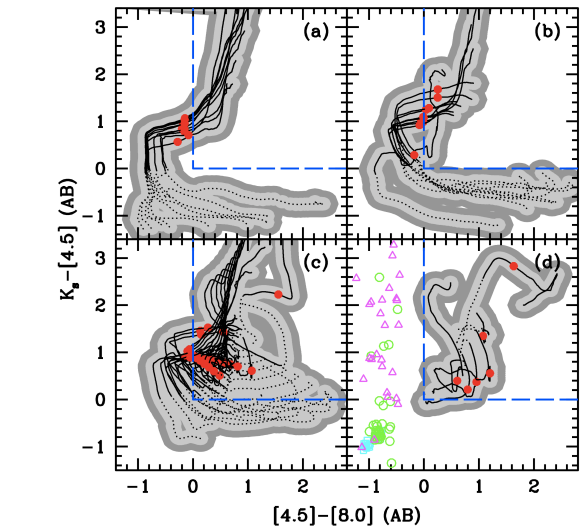}
\caption{The proposed KI criterion. Symbols and panel definition as in Figure~\ref{fig:c1324}.\label{fig:ki}}
\end{figure}

A particularly relevant combination of colors is $K-[4.5]$ versus $[4.5]-[8.0]$ (Figure~\ref{fig:ki}). This $K$+IRAC combination, henceforth called the KI criterion, is defined by the following simple conditions: \[ K-\left[4.5\right]>0 \] and \[ \left[4.5\right]-\left[8.0\right]>0 \] These color cuts result from a balance between rejecting the non-AGN 10--20\% photometric scatter regions at $z\lesssim2.5$ and encompassing the pure-AGN color tracks and respective photometric scatter regions. Using a wide variety of templates and taking into account photometric errors we aim to limit possible selection effects resulting from wedges created \emph{a posteriori}, for example, by using a sample of AGN spectroscopically confirmed such as in L07 and S05.

Similar to the L07, S05, and D12 criteria, the normal galaxy color tracks contaminate the KI selection region at $z\gtrsim2.5$. However, the KI contamination by normal galaxies at $z\lesssim2.5$ appears significantly reduced in comparison to L07 and S05, with no negative effect on the ability to select AGN (i.e., completeness). By avoiding the 10--20\% error scatter region (light- and dark-grey regions) from starburst templates, the KI boundaries have improved efficiency on faint source classification. Another conceptual improvement of KI is the unbounded upper right AGN region. This avoids the loss of heavily obscured AGN (with extremely red colors). This is in contrast to the wedges defined in S05, for example, where the Torus template moves out of the selection region at the highest redshifts ($z\gtrsim4$).

One can also note the usefulness of the simple $K-[4.5]$ color in excluding low redshift normal galaxies: the condition $K-[4.5]>0$ is able to reject a large fraction of the $z<1$ non-AGN galaxies. Such a property also makes this simple color-cut of great use for the study of AGN and star-formation co-evolution in the latter half of cosmic history.

\begin{figure}
\plotone{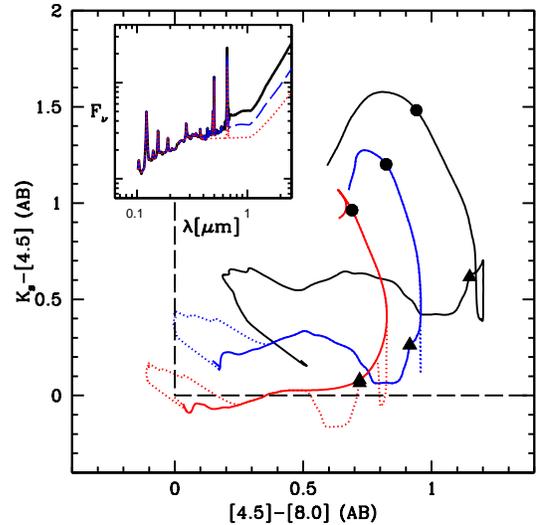}
\caption{The effects of considering prominent lines in type-1 QSO SED models. The small inset shows three such SEDs originally from \citet{Polletta07}: two updated versions of QSO1 (dashed blue line, now with an H$\alpha$ line) and BQSO1 (red dotted, now with both H$\alpha$ and OIII lines), and the TQSO1 template (solid black, which already includes H$\alpha$ and OIII lines). On the main panel the original SED model color tracks are shown as solid lines (from top to bottom: TQSO1, QSO1, and BQSO1), whereas the inclusion of strong emission lines produces the deviations given by the dotted segments. Circles and triangles show $z=1$ and $z=3$, respectively. The tracks extend from $z=0$ to $z=7$.\label{fig:lines}}
\end{figure}

Not all the templates in Table~\ref{tab:temp} take into account prominent emission lines. These may affect the photometry and produce some degree of scatter in the color-color tracks. This is visible in Figure~\ref{fig:lines} where updated versions (highlighted as dotted lines) of the QSO1 and BQSO1 templates \citep{Polletta07} are considered, now with both H$\alpha$ and OIII lines included \citep[visible in the TQSO1 template,][]{Polletta07}. Although in specific redshift intervals (when a certain emission line is redshifted into a given filter), AGN sources with smaller AGN contribution in the IR (like BQSO1) fall out of the AGN region of the KI criterion, the bulk of the AGN population is expected to remain inside the KI boundaries.

\subsection{Extending to high redshifts: the KIM criterion} \label{sec:imhighz}

One serious problem for all criteria investigated so far is the contamination by normal galaxies at $z\gtrsim2.5$. All four (L07, S05, D12, and KI) fail to disentangle AGN dominated systems from normal galaxies at those redshifts. To avoid this problem, longer wavelength observed bands should be considered, allowing the observation of the rest-frame $\lambda>1.6\,\mu$m regime, where AGN emission may dominate, at high redshifts. For this purpose, we extend our study to include the MIPS-$24\,{\mu}m$ band \citep[see also][]{Sajina05}.

The use of this waveband for AGN selection has additional shortcomings. Normal galaxies show a wide IRAC-MIPS color range (as a result of different PAH and dust emissions between galaxies), which is further increased by redshift (the PAH features are redshifted out of the 8.0\,$\mu$m band at low-$z$, but redshifted into the 24\,$\mu$m band at $z\sim2$). This results in a considerable color overlap with AGNs, limiting the usefulness of a single IRAC-MIPS color to separate both populations \citep{Lacy04,Hatziminaoglou05,Cardamone08}. Nevertheless, some authors have used the MIPS-$24\,{\mu}m$ band for unique, extreme objects (e.g., the IR-excess technique, Section~\ref{sec:irxs}) or in single, unconventional situations. For instance, while \citet{Garn09} use $[8.0]-[24]$ against $[5.8]-[8.0]$ for a $z\sim0.8$ sample to identify those sources showing AGN activity, \citet{Treister06} and \citet{Messias10} use a single $[8.0]-[24]$ color cut at, respectively, $z\sim2$ and $z>2.5$ for the same purpose\footnote{\citet{Ivison04} and \citet{Pope08} also explore $[8.0]-[24]$ against $[4.5]-[8.0]$ to distinguish AGN from normal galaxies, but, in those works, only the $[4.5]-[8.0]$ color is effectively used for that purpose.}.

Colors involving the 24\,$\mu$m band are also usually avoided due to the large wavelength gap between this band and other commonly available MIR bands (usually the \textit{Spitzer}-IRAC bands). At high-$z$ (e.g., $z\sim3$), however, the sampled rest-frame wavebands (2 and 6\,$\mu$m corresponding to observed 8 and 24\,$\mu$m, respectively) are not much more separated than the 3.6 and 8.0\,${\mu}m$ IRAC bands for nearby galaxies. A different issue is the lower sensitivity and larger point spread function of the MIPS 24\,$\mu$m images when compared with those obtained with the IRAC channels, which affects the accuracy of color measurements using this longer wavelength band. Nevertheless, efficient photometry algorithms have been built to overcome these blending problems \citep{Wuyts08,Santini09}.

\begin{figure}
\plotone{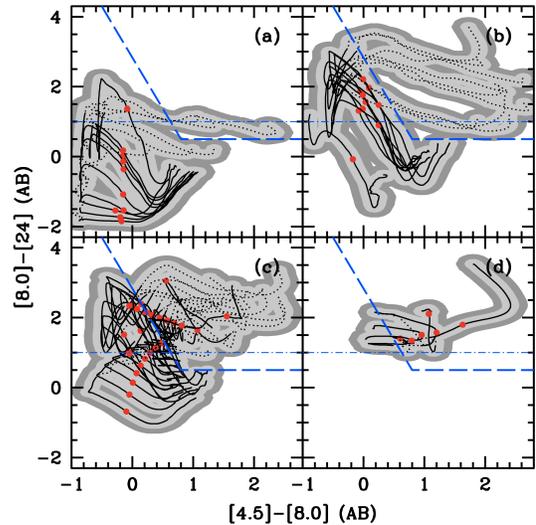}
\caption{The IRAC-MIPS color-color space and the proposed criterion. Symbols and panel definition as in Figure~\ref{fig:c1324}. The dot-dashed horizontal blue line (at $[8.0]-[24]=1$) refers to a simpler criterion valid at $z\gtrsim3$, as detailed in the text and illustrated in Figure~\ref{fig:z824}.\label{fig:c24m24}}
\end{figure}

Figure~\ref{fig:c24m24} illustrates a color-color AGN diagnostic proposed for $z\gtrsim1$, with its boundaries avoiding a 10\% photometric error scatter from normal galaxies. The criterion is defined by the following IRAC-MIPS (IM) conditions: \[\left[8.0\right]-\left[24\right] > -2.9\times(\left[4.5\right]-\left[8.0\right])+2.8\] and \[\left[8.0\right]-\left[24\right] > 0.5\]

One can see that beyond $z\sim1$, AGN (lower panels) and normal galaxies (upper panels) occupy essentially different regions in the [8.0]-[24] versus [4.5]-[8.0] space. This is of great interest for the characterization of high redshift galaxy populations, such as Lyman Break Galaxies (LBGs) and equivalents at $z\gtrsim2$ \citep{Steidel03,Steidel04,Adelberger04}, or spectroscopically confirmed high-$z$ galaxies. We note, nevertheless, the incompleteness toward type-1 QSOs at $z\gtrsim4-5$\footnote{However, type-1 QSOs with the largest optical-to-IR flux ratios migrate back in the AGN region at $z\gtrsim6$.} or type-2 QSOs with strong H$_\alpha$ emission affecting the $4.5\,\mu$m band at $z\gtrsim4.5$ (Figure~\ref{fig:lines}). Note, however, that Mrk231 and torus-type AGN remain inside the region for the full redshift range studied ($0<z<7$).

At low redshifts ($z\lesssim1$), degeneracy exists in this IM color-color space, with AGN and normal galaxies occupying the same color-color region. A rejection of low-redshift ($z<1$) normal/star forming galaxies would, however, remove this overlap, allowing for a powerful AGN-selection criterion to be built. This can be achieved, as noted in the previous section, by using the $K-[4.5]>0$ color cut (Figure~\ref{fig:ki}). This allows the rejection of a large fraction of the $z\lesssim1$ non-AGN galaxies, with AGN and hybrid galaxies in this redshift range remaining mostly unaffected.

The IRAC-MIPS (IM) conditions presented above, when considered together with the $K-[4.5]>0$ cut, which implements the rejection of $z<1$ normal galaxies, define what we will henceforth call the KIM (K+IRAC+MIPS) criterion.

\begin{figure}
\plotone{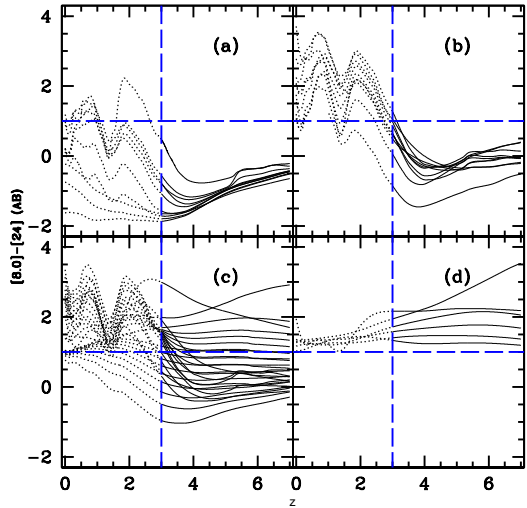}
\caption{The [8.0]-[24] color evolution with redshift. Panel definition as in Figure~\ref{fig:c1324}. The horizontal line shows $[8.0]-[24]=1$, while the vertical one indicates $z=3$. At $z>3$, only AGN dominated
galaxies show $[8.0]-[24]>1$ colors.\label{fig:z824}}
\end{figure}

We further note from Figure~\ref{fig:c24m24} that for $z\gtrsim3$, essentially all SEDs with $[8.0]-[24]>1$ are dominated by AGN emission. This happens because in this redshift range, the PAH features dominating a normal galaxy SED are redshifted out of the MIPS$_{\rm 24\mu{m}}$ filter. Figure~\ref{fig:z824} details this behavior, clearly showing that stellar dominated galaxies at $z\gtrsim3$ show $[8.0]-[24]<1$ colors, as described by \citet{Messias10}. Such a color cut, if possible to apply (one has to know the source redshift), allows the recovery of H$\alpha$-emitting and/or rest-frame UV/optical blue QSOs expected to leave the IM region at $z\gtrsim4-5$ (Figure~\ref{fig:c24m24}).

We acknowledge that by aiming at such a broad redshift range ($0<z\lesssim7$) KIM has its shortcomings. For instance, if the goal is to characterize sources at $z\lesssim1$, the use of $K-[4.5]$ allows a reliable selection of AGN-hosts avoiding the need for longer wavelength bands. On the other hand, at $z\gtrsim3$, the $[8.0]-[24]$ color is enough to identify AGN dominated sources, avoiding the loss of sources undetected at shorter wavelengths. Finally, the $[4.5]-[8.0]$ colour is best at intermediate redshifts ($1\lesssim{z}\lesssim2.5$).

However, a redshift estimate is frequently not available (incomplete spectroscopic sample coverage) nor reliable (photometric redshifts). Particularly for AGN-dominated sources, where a power-law SED is observed, the lack of spectral features (such as spectrum breaks) makes the photometric redshift solutions degenerate. In order to avoid such limitations, while still allowing AGN selection up to $z\sim7$, the KIM criterion may be valuable.

\section{Testing against multi-wavelength AGN samples} \label{sec:ctrlsp}

In the previous section we have proposed:

\begin{itemize}
\item[-] $K-[4.5]$ as an useful color for the efficient segregation of the galaxy population into AGN-dominated and normal SEDs at $z\lesssim1$;
\item[-] KI criterion as an alternative to L07, S05, and D12; and
\item[-] KIM (a 4 band, 3 color criterion), as a diagnostic which, according to the color tracks of the templates used, enables the selection of AGN sources at $0<z\lesssim7$ with little contamination by normal galaxies.
\end{itemize}

This is of great interest as it allows one to trace AGN activity since the epoch of reionization to the current time. The usefulness of these criteria can only be evaluated, however, by pursuing a test with well characterized control samples. Using various samples for which normal-galaxy and AGN classifications are obtained via other spectral regimes (from X-rays to radio), we can obtain some estimate of the efficiency of the new proposed diagnostics in comparison with commonly used ones.

Because an ideal sample, where \emph{all} the sources are \emph{reliably} split between AGN and non-AGN hosts, does not exist, we will perform these tests using five control samples. First, we use a sample of galaxies from the \emph{Chandra} Deep Field South \citep[CDFS,][]{Giacconi01} and another from the Cosmic Evolution Survey \citep[COSMOS,][]{Scoville07}, both with available AGN/non-AGN classification from X-rays and/or optical spectroscopy. Second, we assemble samples of IR-excess (IRxs) sources \citep{Dey08,Fiore08,Polletta08} found in CDFS and COSMOS fields. The QSO sample from the Sloan Digital Sky Survey \citep[SDSS,][]{Schneider10}, reaching $z\sim6$, is also considered for the testing, as well as the High-\textit{z} Radio Galaxy (H\textit{z}RG) sample from \citet{Seymour07}. The first two samples (X-ray/optical selected in CDFS and COSMOS) allow for an indication of the completeness and non-AGN contamination of the IR AGN selection criteria, while the AGN samples (IRxs sources, SDSS QSOs and H\textit{z}RGs) will allow for independent measures of their completeness up to the highest redshifts.

In the following subsections, Completeness ($\mathcal{C}$) is defined as the fraction of the AGN population which a given IR criterion selects (AGN$_{SEL}$/AGN$_{TOT}$), while Reliability ($\mathcal{R}$) refers to the fraction of the IR sources selected by a given criterion which are ``true'' AGN ($\rm{AGN_{SEL}/N_{SEL}}$, where $\rm{N_{SEL}=AGN_{SEL}+non}$-$\rm{AGN_{SEL}}$).

\subsection{The CDFS and COSMOS samples} \label{sec:goodssample}

We have selected 2287 galaxies from MUSIC/CDFS catalog \citep{Grazian06,Santini09} and 7217 from COSMOS \citep{Ilbert09} with an X-ray classification (based on source luminosity and spectral slope) and/or a good quality\footnote{Spectra flagged as 0 (very good) or 1 (good) in the MUSIC catalog, and estimates with $>90\%$ probability of being correct in the $z$COSMOS catalog.} optical spectroscopic classification. Whenever a spectroscopic redshift was not available, the photometric estimates by \citet{Luo10} and \citet{Salvato11} were adopted.

The IR data used for the MUSIC catalog comes from \citet{Vandame02} and Dickinson et al. (in prep.), and that for COSMOS comes from \citet{Sanders07}, \citet{Lefloch09}, and \citet{McCracken10}. Regarding the X-rays, the 2\,Ms \textit{Chandra} Deep Field South \citep[CDF-S,][]{Luo08} data was used, as well as the XMM data in COSMOS \citep{Cappelluti09,Brusa10}. The X-ray AGN classification is similar to that of \citet{Szokoly04}. There, the X-ray luminosity and hardness-ratio (HR) are used to identify the AGN population. The HR is a measure of the source obscuration and is defined as HR$\equiv$(H-S)/(H+S) with H and S being, respectively, the net counts in the hard, 2--8\,keV, and soft, 0.5--2\,keV, X-ray bands. However, this ratio becomes degenerated with redshift (\citealt{Eckart06} and \citealt{Messias10}, but also \citealt{Alexander05} and \citealt{Luo10}), and count measurements depend on array efficiency between telescopes. Hence we compute for each source the respective column densities ($\rm{N_H}$) using the Portable, Interactive Multi-Mission Simulator\footnote{http://heasarc.nasa.gov/docs/software/tools/pimms.html} (PIMMS, version 3.9k). The soft-band/full-band (SB/FB) and hard-band/full-band (HB/FB) flux ratios\footnote{The use of ratios based on FB flux instead of the commonly used SB/HB flux ratios, allows for an estimate of $\rm{N_H}$ when the source is detected in the FB but no detection is achieved in either the SB or HB.} were estimated for a range of column densities ($19<\log(\rm{N_H[{\rm cm}^{-2}]})<25$, with steps of $\log(\rm{N_H[{\rm cm}^{-2}]})=0.01$), and redshifts ($0<z<7$, with steps of $z=0.01$), considering a fixed photon index, $\Gamma=1.8$ \citep{Tozzi06}. The comparison with the observed values results in the estimate of $\rm{N_H}$. However, it is known that this procedure systematically overestimates the intrinsic $\rm{N_H}$. The Appendix discusses this bias and how we attempted to correct it. The final $\rm{N_H}$ value is then used to derive an intrinsic X-ray luminosity. The HR constraint used by \citet{Szokoly04} ($\rm{HR}=-0.2$) is equivalent to $\log(\rm{N_H[{\rm cm}^{-2}]})=22$ at $z\sim0$, and this is the value considered throughout the whole redshift range.

Hence, an X-ray AGN is considered to have $\rm{L_X^{int}}>10^{41}\,\rm{erg\,s^{-1}}$ and $\rm{N_H}>10^{22}\,{\rm cm}^{-2}$, or $\rm{L_X^{int}}>10^{42}\,\rm{erg\,s^{-1}}$. The remaining X-ray detections are regarded as non-AGN sources. The intrinsic X-ray (0.5--10\,keV) luminosities are estimated as: \[ \rm{L_X^{int}} = 4 \pi\,d^{2}_{L}\,f_{X}^{int}\,(1+z)^{\Gamma-2}\,\rm{erg}\,\rm{s}^{-1} \] where $\rm{f_{X}^{int}}$ is the obscuration-corrected X-ray flux in the 0.5--10 keV band, and $\Gamma$ is the observed photon index (when $\log(\rm{N_H[{\rm cm}^{-2}]})\leq19$) or $\Gamma=1.8$ (when $\log(\rm{N_H[{\rm cm}^{-2}]})>19$). In CDFS, the 0.5--8~keV luminosities are derived using \citet{Luo08} catalogued 0.5--8~keV fluxes and converted to 0.5--10~keV considering the adopted $\Gamma$. For simplicity, the luminosity `int' label is dropped from now on, as we will always be referring to intrinsic luminosities, unless otherwise stated.

Regarding the spectroscopic sample, the AGN sources are those which display broad line features or high-ionization narrow emission lines characteristic of AGN activity (BLAGN or NLAGN, respectively). The remaining sources with a spectroscopic classification are regarded as part of the non-AGN population (e.g., SF galaxies, stars). The NLAGN classification comes from the MUSIC catalog in CDFS, and from \citet{Bongiorno10} in COSMOS.

In both the CDFS and COSMOS final AGN samples, most sources have an X-ray AGN classification (82\% and 80\%, respectively), and a significant fraction has a spectroscopic AGN classification (21\% in CDFS and 55\% in COSMOS, the difference resulting from different spectroscopic completenesses between the two surveys).

\subsubsection{KI/KIM efficiency in CDFS} \label{sec:kikimgoodss}

\begin{figure}
\plotone{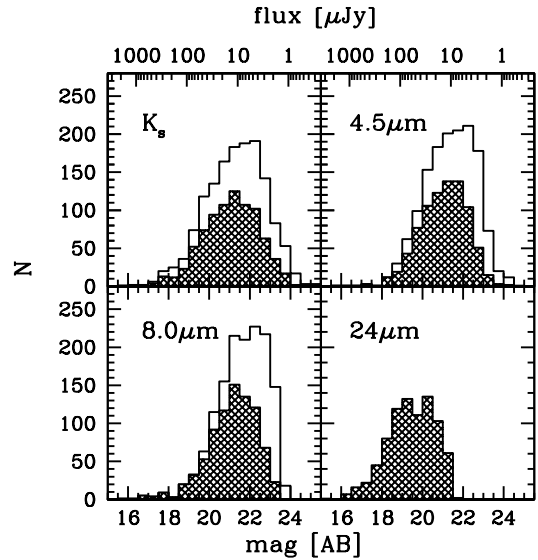}
\caption{The magnitude distribution of the final CDFS sample with reliable photometry in K-IRAC bands (open histogram) and K-IRAC-MIPS$_{24\mu{\rm m}}$ (hatched histogram). Each panel refers to the magnitude distribution in the following bands: $K_s$ (upper left), 4.5$\,\mu$m (upper right), 8.0$\,\mu$m (lower left), and 24$\,\mu$m (lower right). Note that in the latter both histograms coincide.\label{fig:magdist}}
\end{figure}

For consistency, we only consider sources having photometry estimates with a flux error smaller than a third of the flux measurement (equivalent to an error in magnitude smaller than 0.36) in all $K_s$-IRAC bands when testing L07, S05, D12, and KI. This requirement will remove many of the fainter objects, but the final sample is still among the deepest ever used to test these IR criteria. The magnitude distribution of the final sample considered is shown in Figure~\ref{fig:magdist} (see open histograms). Among the 1440 sources composing the final sample, 166 (12\%) are classified as AGN hosts (160 in X-rays, 38 through spectroscopy). The sample is further separated into redshift ranges: $0\leq{z}<1$, $1\leq{z}<2.5$, $2.5\leq{z}<4$. The adopted threshold at $z=2.5$ is based on the discussion in section~\ref{sec:ki}. This results in 801, 535, and 94 sources with $K_s$-IRAC photometry at $0\leq{z}<1$, $1\leq{z}<2.5$, and $2.5\leq{z}<4$, respectively.

When testing KIM we also require reliable 24\,$\mu$m photometry (hatched histograms in Figure~\ref{fig:magdist}). However, this requirement restricts the sample to the brightest sources, unavoidably increasing the probability of finding AGN dominated sources \citep{Brand06,Treister06,Donley08}. Although it can be considered a constraint by itself (which improves the criterion efficiency), this results in an unfair comparison of the actual color-color regions. Hence, when comparing KIM to L07, S05, and KI, we consider the sample of 834 sources (460/324/47 at $0\leq{z}<1$, $1\leq{z}<2.5$, and $2.5\leq{z}<4$, respectively) with reliable $K_s$-IRAC-MIPS$_{24\,\mu{\rm m}}$ photometry, of which 139 (17\%) are classified as AGN hosts.

Tables~\ref{tab:xrspcs}, \ref{tab:xrspcs125}, and \ref{tab:xrspcs254} summarize the final statistics for the application of each of the IR criteria to the CDFS control sample at different redshift ranges. In order to assess the significance of the results, two types of error are presented: Poisson and photometric. The latter is computed by scattering the source colors between the $\pm1\sigma$ magnitude errors (which we limited to $<1/3$ of the flux value) and evaluate how completeness ($\mathcal{C}$) and reliability ($\mathcal{R}$) vary as a result. However, there is a covariance between observational photometric and Poisson sampling errors, since the samples with the smallest photometric errors are the smallest in number, leading in turn to large sampling errors. Since our approach is based around testing colour-space boundaries that have already accounted for generous photometric variation around the SED templates in their definition, the current analysis is primarily limited by sampling (Poisson) errors. In general, the Poisson errors are larger than those that would be inferred from the photometric errors alone, so this approach is also conservative in the sense of erring in favour of the larger uncertainty. The discussion and results presented below are consequently based on a consideration of the Poisson errors associated with the sample sizes being used in the analysis of Completeness and Reliability.

\begin{deluxetable}{ccrrcc}
\tabletypesize{\scriptsize}
\tablecaption{CDFS X-ray and Spectroscopic $0\leq{z}<1$ control sample test.\label{tab:xrspcs}}
\tablewidth{0pt}
\tablehead{
\colhead{Sample} & \colhead{Criterion} & \colhead{N$_{SEL}$\tablenotemark{a}} & \colhead{AGN\tablenotemark{b}} & \colhead{{$\cal C$}\tablenotemark{c}} & \colhead{{$\cal{R}$}\tablenotemark{c}}
}
\startdata
\smallskip
$K$+IRAC & [none]	& 801	& 42	& \ldots		& (5$\pm$1) \\ 
\smallskip
& L07		& 105	& 21	& 50(13,$_{2}^{2}$)	& 20(5,$_{8}^{9}$) \\ 
\smallskip
& S05		& 26	& 12	& 29(9,$_{5}^{5}$)	& 46(16,$_{13}^{15}$) \\ 
\smallskip
& D12		& 5	& 4	& 10(5,$_{2}^{2}$)	& 80(54,$_{20}^{3}$) \\ 
\smallskip
& KI		& 24	& 12	& 29(9,$_{0}^{0}$)	& 50(18,$_{2}^{13}$) \\ 
\hline
\smallskip
$K$+IRAC+ & [none]	& 460	& 37	& \ldots		& (8$\pm$1) \\ 
\smallskip
MIPS$_{24\,\mu\rm{m}}$ & L07		& 76	& 18	& 49(14,$_{0}^{3}$)	& 24(6,$_{7}^{6}$) \\ 
\smallskip
& S05		& 20	& 11	& 30(10,$_{5}^{3}$)	& 55(21,$_{16}^{5}$) \\ 
\smallskip
& D12		& 5	& 4	& 11(6,$_{3}^{0}$)	& 80(54,$_{20}^{0}$) \\ 
\smallskip
& KI		& 17	& 10	& 27(10,$_{0}^{0}$)	& 59(23,$_{3}^{8}$) \\ 
\smallskip
& KIM		& 10	& 7	& 19(8,$_{0}^{0}$)	& 70(34,$_{20}^{0}$) \\ 
\enddata
\tablecomments{This table is restricted to the $0\leq{z}<1$ CDFS sample. While in the upper group of rows reliable photometry is required --- a magnitude error below 0.36 --- in $K$+IRAC bands, in the lower group of rows we also require reliable 24\,${\mu}m$ photometry. The first row in each group refers to the total number of sources with reliable $K$+IRAC (upper group) and $K$+IRAC+24\,$\mu$m (bottom group) photometry. For reference, the value in parenthesis (with the corresponding Poisson error) in the $\cal{R}$ column gives the overall fraction of identified AGN hosts, equivalent to the $\cal{R}$ of a criterion selecting all sources in the field with reliable photometry in the considered bands.}
\tablenotetext{a}{Number of sources selected by a given criterion with an AGN/non-AGN classification from X-rays and/or spectroscopy.}
\tablenotetext{b}{Number of selected sources with an AGN classification, from either the X-rays or optical spectroscopy.}
\tablenotetext{c}{Completeness ($\cal C$) calculated as AGN$_{\rm{SEL}}$/AGN$_{\rm{TOT}}$. Reliability ($\cal R$) calculated as AGN$_{\rm{SEL}}$/N$_{\rm{SEL}}$. Numbers in parenthesis refer, respectively, to the Poisson error and the photometric upper and lower error bars.}
\end{deluxetable}

\begin{deluxetable}{ccrrcc}
\tabletypesize{\scriptsize}
\tablecaption{CDFS X-ray and Spectroscopic $1\leq{z}<2.5$ control sample test.\label{tab:xrspcs125}}
\tablewidth{0pt}
\tablehead{
\colhead{Sample} & \colhead{Criterion} & \colhead{N$_{SEL}$} & \colhead{AGN} & \colhead{{$\cal C$}} & \colhead{{$\cal{R}$}}
}
\startdata
\smallskip
$K$+IRAC & [none]	& 535	& 80	& \ldots		& (15$\pm$2) \\ 
\smallskip
& L07		& 170	& 50	& 62(11,$_{2}^{6}$)	& 29(5,$_{8}^{12}$) \\ 
\smallskip
& S05		& 104	& 28	& 35(8,$_{7}^{6}$)	& 27(6,$_{13}^{22}$) \\ 
\smallskip
& D12		& 18	& 15	& 19(5,$_{4}^{0}$)	& 83(29,$_{20}^{17}$) \\ 
\smallskip
& KI		& 60	& 32	& 40(8,$_{4}^{1}$)	& 53(12,$_{21}^{14}$) \\ 
\hline
\smallskip
$K$+IRAC+ & [none]	& 324	& 61	& \ldots		& (19$\pm$3) \\ 
\smallskip
MIPS$_{24\,\mu\rm{m}}$ & L07		& 110	& 39	& 64(13,$_{2}^{5}$)	& 35(7,$_{6}^{9}$) \\ 
\smallskip
& S05		& 70	& 25	& 41(10,$_{7}^{5}$)	& 36(8,$_{15}^{17}$) \\ 
\smallskip
& D12		& 15	& 14	& 23(7,$_{5}^{0}$)	& 93(35,$_{9}^{7}$) \\ 
\smallskip
& KI		& 40	& 26	& 43(10,$_{2}^{2}$)	& 65(16,$_{16}^{8}$) \\ 
\smallskip
& KIM		& 25	& 16	& 26(7,$_{3}^{3}$)	& 64(20,$_{14}^{14}$) \\ 
\enddata
\tablecomments{This table is restricted to the $1\leq{z}<2.5$ CDFS sample. Table structure and column definition as in Table~\ref{tab:xrspcs}.}
\end{deluxetable}

\begin{deluxetable}{ccrrcc}
\tabletypesize{\scriptsize}
\tablecaption{CDFS X-ray and Spectroscopic $2.5\leq{z}<4$ control sample test.\label{tab:xrspcs254}}
\tablewidth{0pt}
\tablehead{
\colhead{Sample} & \colhead{Criterion} & \colhead{N$_{SEL}$} & \colhead{AGN} & \colhead{{$\cal C$}} & \colhead{{$\cal{R}$}}
}
\startdata
\smallskip
$K$+IRAC & [none]	& 94	& 40	& \ldots		& (43$\pm$8) \\ 
\smallskip
& L07		& 93	& 40	& 100(22,$_{5}^{0}$)	& 43(8,$_{2}^{3}$) \\ 
\smallskip
& S05		& 54	& 29	& 72(18,$_{20}^{10}$)	& 54(12,$_{18}^{23}$) \\ 
\smallskip
& D12		& 29	& 20	& 50(14,$_{12}^{18}$)	& 69(20,$_{27}^{27}$) \\ 
\smallskip
& KI		& 73	& 36	& 90(21,$_{5}^{3}$)	& 49(10,$_{7}^{8}$) \\ 
\hline
\smallskip
$K$+IRAC+ & [none]	& 47	& 33	& \ldots		& (70$\pm$16) \\ 
\smallskip
MIPS$_{24\,\mu\rm{m}}$ & L07		& 47	& 33	& 100(25,$_{6}^{0}$)	& 70(16,$_{1}^{2}$) \\ 
\smallskip
& S05		& 32	& 24	& 73(20,$_{15}^{9}$)	& 75(20,$_{14}^{7}$) \\ 
\smallskip
& D12		& 20	& 17	& 52(15,$_{9}^{18}$)	& 85(28,$_{11}^{15}$) \\ 
\smallskip
& KI		& 41	& 30	& 91(23,$_{3}^{0}$)	& 73(18,$_{2}^{4}$) \\ 
\smallskip
& KIM		& 24	& 20	& 61(17,$_{9}^{12}$)	& 83(25,$_{15}^{9}$) \\ 
\enddata
\tablecomments{This table is restricted to the $2.5\leq{z}<4$ CDFS sample. Table structure and column definition as in Table~\ref{tab:xrspcs}.}
\end{deluxetable}

In the low redshift bin ($0\leq{z}<1$), L07 is the most complete (selecting almost 50\% of the AGN sample), but the least reliable as well, with less than a fifth of the sample being confirmed AGN hosts. This is understood given the large selection region set by the color constraints of L07 criterion, which despite selecting many AGN hosts (high $\mathcal{C}$), also selects many non-AGN sources (low $\mathcal{R}$). S05, KI, and KIM present almost comparable efficiencies, selecting around a quarter of the AGN population, and with 50--60\% reliabilities. D12 retrieves the lowest $\mathcal{C}$ values (10\%, comparable with KIM, see also Section~\ref{cosmosctrl}), but at the same time the highest $\mathcal{R}$ values (80\%, although with large Poisson errors).

At $1\leq{z}<2.5$, L07 is still the criterion that provides the most complete AGN sample, but here it presents an $\mathcal{R}$ value of $\sim30\%$, comparable to that of S05. Both KI and KIM present improved $\mathcal{R}$ levels, with 50--60\% of the sample being confirmed as AGN hosts. The improvement relative to the L07 and S05 criteria in this redshift range appears to be due to the $[4.5]-[8.0]$ color cut used. The D12 criterion is again the criterion that presents the lowest $\mathcal{C}$ (20\%) and the highest $\mathcal{R}$ (80--90\%). The D12 criterion is affected by small sample statistics, but it always recovers the highest $\mathcal{R}$ values, supporting the scenario that it is in fact the most reliable of all the tested criteria.

At high-$z$ ($2.5\leq{z}<4$), the fraction of ``true" AGN hosts is already high (43\%, increasing to 70\% when restricting to the MIPS$_{24\,\mu{\rm m}}$ detected sample). As a result of the small number of sources (driven by the necessary quality constraints), no conclusion can be drawn regarding $\mathcal{R}$ from this sample at such high redshifts, given that all criteria provide comparable $\mathcal{R}$ results within the uncertainties. However, differences in completeness exist. D12 and KIM show the lowest values ($\mathcal{C}\sim50-60\%$), KI and L07 the highest ($\mathcal{C}\sim90-100\%$), while S05 gives intermediate values ($\mathcal{C}\sim70\%$).

One should also note the tendency for S05 and KIM to present the largest range between lower and upper photometric limits. While in KIM this relates to the use of the less sensitive MIPS$_{24\,\mu{\rm m}}$ band, the larger photometric error ranges in S05 are due to the use of bands close in wavelength (Figure~\ref{fig:c1324}), instead of bands widely separated in wavelength as in L07, KI and KIM. The large photometric errors found for the D12 criterion are probably related to the small sample statistics (see COSMOS results below, where these errors are close to zero).

\begin{figure}
\plotone{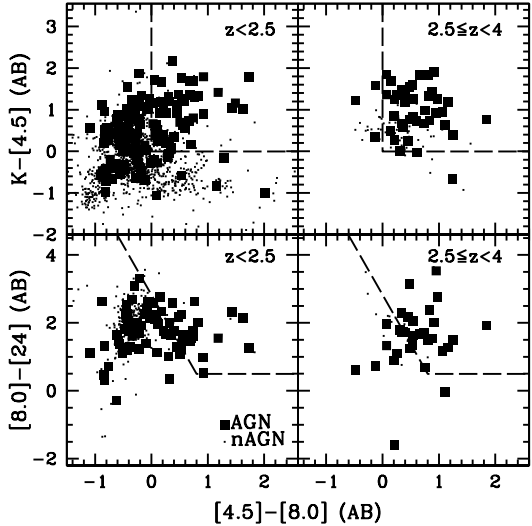}
\caption{The CDFS sources in KI (upper panels) and KIM (lower panels) color-color spaces, divided into low-$z$ ($z<2.5$, left panels) and high-$z$ ($2.5\leq{z}<4$, right panels) groups. Squares represent
AGN hosts (as classified from X-rays or optical spectroscopy), while dots highlight non-AGN sources. The dashed lines in the upper panels refer to the KI criterion, while the dashed lines in the lower panels refer to the IM region proposed in Section~\ref{sec:imhighz}. All sources displayed in the lower panels have $K-[4.5]>0$, as required by the KIM criterion.\label{fig:goodsamp}}
\end{figure}

Finally, Figure~\ref{fig:goodsamp} details the application to the CDFS data of KI (upper panels) and KIM (lower panels), separated into redshift bins ($z<2.5$ and $2.5\leq{z}<4$). For this exercise, we have required reliable photometry in the bands needed for KI or KIM. This figure shows that the KI and KIM criteria still miss a significant fraction (around two thirds) of the AGN population at $z<2.5$. The high redshift panels show that most (around three quarters) of the overall population indeed falls inside the KI selection region (in agreement with Figure~\ref{fig:ki}). The low number of blue $[8.0]-[24]$ sources is mainly a result of the photometric quality constraints applied to the sample. Knowing that the MIPS$_{24\,\mu{\rm m}}$ band is also the less sensitive waveband, at high redshifts, where sources tend to be fainter, blue $[8.0]-[24]$ sources will likely not be detected in the MIPS$_{24\,\mu{\rm m}}$ band and consequently will be absent from this exercise.

When no redshift estimate is available, the previous analysis, which separates the sample into redshift bins, would not be possible. Table~\ref{tab:xrspcsall} shows the results for the overall CDFS sample with no redshift constraint. The conclusions remain the same: KI and KIM are complete to $\sim$50\% and $\sim$35\% levels, respectively, and both show improved reliability ($\mathcal{R}\sim50-70\%$) relative to L07 and S05 (KI and S05 error bars marginally overlap); and D12 is less complete (with values comparable to KIM, $\mathcal{C}=25-30\%$), but the most reliable ($\mathcal{R}=76-88\%$).

\begin{deluxetable}{ccrrcc}
\tabletypesize{\scriptsize}
\tablecaption{CDFS X-ray and Spectroscopic control sample test.\label{tab:xrspcsall}}
\tablewidth{0pt}
\tablehead{
\colhead{Sample} & \colhead{Criterion} & \colhead{N$_{SEL}$} & \colhead{AGN} & \colhead{{$\cal C$}} & \colhead{{$\cal{R}$}}
}
\startdata
\smallskip
$K$+IRAC & [none]	& 1440	& 166	& \ldots		& (12$\pm$1) \\ 
\smallskip
& L07		& 377	& 115	& 69(8,$_{3}^{4}$)	& 31(3,$_{8}^{9}$) \\ 
\smallskip
& S05		& 188	& 73	& 44(6,$_{10}^{7}$)	& 39(5,$_{17}^{22}$) \\ 
\smallskip
& D12		& 55	& 42	& 25(4,$_{6}^{5}$)	& 76(16,$_{25}^{20}$) \\ 
\smallskip
& KI		& 166	& 84	& 51(7,$_{3}^{1}$)	& 51(7,$_{12}^{11}$) \\ 
\hline
\smallskip
$K$+IRAC+ & [none]	& 834	& 134	& \ldots		& (16$\pm$1) \\ 
\smallskip
MIPS$_{24\,\mu\rm{m}}$ & L07		& 236	& 93	& 69(9,$_{2}^{3}$)	& 39(5,$_{8}^{7}$) \\ 
\smallskip
& S05		& 125	& 63	& 47(7,$_{9}^{5}$)	& 50(8,$_{17}^{14}$) \\ 
\smallskip
& D12		& 42	& 37	& 28(5,$_{5}^{4}$)	& 88(20,$_{11}^{10}$) \\ 
\smallskip
& KI		& 101	& 69	& 51(8,$_{1}^{1}$)	& 68(11,$_{9}^{6}$) \\ 
\smallskip
& KIM		& 62	& 46	& 34(6,$_{4}^{4}$)	& 74(14,$_{16}^{10}$) \\ 
\enddata
\tablecomments{No redshift cut is considered in this table. Table structure and column definition as in Table~\ref{tab:xrspcs}.}
\end{deluxetable}

\subsubsection{KI/KIM efficiency in COSMOS} \label{cosmosctrl}

The same comparison is now performed for COSMOS. No redshift segregation is applied as there is no classified SF system at $z\gtrsim1.6$ in this COSMOS sample. Among the 7216 sources with either a spectral or X-ray classification and adequate $K$-IRAC photometry, 1423 are flagged as AGN hosts by X-ray and spectroscopy criteria. There are 2641 sources with MIPS$_{24\mu{\rm m}}$ detection (842 AGN hosts). Table~\ref{tab:Cxrspcs} reports the final statistics on the application of the various diagnostics. Note that 84\% of this COSMOS sample is found at $z<1$, implying that the statistics of this sample will be dominated by those of the $z<1$ population. Hence, it is fair to compare these results with those found for the CDFS sample at $0\leq{z}<1$ (Table~\ref{tab:xrspcs}). The COSMOS sample nevertheless provides a more numerous sample, giving smaller Poisson errors. The L07 criterion remains the most complete ($\mathcal{C}\sim80\%$), but also the least reliable ($\mathcal{C}\sim55-70\%$), as opposed to D12, which shows the highest reliability levels ($\mathcal{R}\sim100\%$). KI marginally retrieves a more reliable AGN sample than S05, but comparable completenesses ($\mathcal{C}\sim60-70\%$ and $\mathcal{R}\sim90\%$). KIM shows a reliability of $\sim90\%$, but is likely the least complete criteria at $z\lesssim1$.

\begin{deluxetable}{ccrrcc}
\tabletypesize{\scriptsize}
\tablecaption{COSMOS X-ray and Spectroscopic control sample test.\label{tab:Cxrspcs}}
\tablewidth{0pt}
\tablehead{
\colhead{Sample} & \colhead{Criterion} & \colhead{N$_{SEL}$} & \colhead{AGN} & \colhead{{$\cal C$}} & \colhead{{$\cal{R}$}}
}
\startdata
\smallskip
$K$+IRAC & [none]	& 7216	& 1423	& \ldots		& (20$\pm$1) \\ 
\smallskip
& L07		& 2038	& 1115	& 78(3,$_{3}^{3}$)	& 55(2,$_{9}^{10}$) \\ 
\smallskip
& S05		& 1102	& 921	& 65(3,$_{7}^{5}$)	& 84(4,$_{13}^{9}$) \\ 
\smallskip
& D12		& 586	& 573	& 40(2,$_{4}^{5}$)	& 98(6,$_{3}^{0}$) \\ 
\smallskip
& KI		& 967	& 881	& 62(3,$_{3}^{4}$)	& 91(4,$_{4}^{3}$) \\ 
\hline
$K$+IRAC+ & [none]	& 2641	& 842	& \ldots		& (32$\pm$1) \\ 
\smallskip
MIPS$_{24\,\mu\rm{m}}$ & L07		& 1087	& 727	& 86(4,$_{2}^{1}$)	& 67(3,$_{5}^{5}$) \\ 
\smallskip
& S05		& 701	& 629	& 75(4,$_{4}^{2}$)	& 90(5,$_{4}^{4}$) \\ 
\smallskip
& D12		& 464	& 454	& 54(3,$_{2}^{3}$)	& 98(6,$_{0}^{0}$) \\ 
\smallskip
& KI		& 643	& 588	& 70(4,$_{1}^{1}$)	& 91(5,$_{2}^{2}$) \\ 
\smallskip
& KIM		& 435	& 398	& 47(3,$_{4}^{3}$)	& 91(6,$_{3}^{2}$) \\ 
\enddata
\tablecomments{Table structure and column definition as in Table~\ref{tab:xrspcs}. No redshift range is adopted as there is no classified SF system at $z\gtrsim1.6$ in the COSMOS sample.}
\end{deluxetable}

It is difficult to directly compare in absolute value the results achieved with the CDFS and COSMOS samples, since many characteristics differ between the two surveys. As an example, by applying the COSMOS (IR and X-rays) flux limits to the CDFS sample, the $\mathcal{C}$ and $\mathcal{R}$ values for the different diagnostics are closer to those of COSMOS. Other issues may contribute to this, such as (a) different photometry extraction methods (ConvPhot by MUSIC team in CDFS and aperture photometry in COSMOS), (b) different spectral coverage depth and procedures for spectral classification, (c) the different photon indices used for the CDFS and COSMOS samples to convert from count rates to X-ray fluxes, (d) differences in relative sensitivity between soft and hard bands of the \textit{Chandra Space Telescope} (in CDFS) and XMM-\textit{Newton} (in COSMOS), and (e) cosmic variance. We stress, however, that the results on the relative efficiency between the criteria are not contradictory between the two fields.

\subsection{IR-excess sources} \label{sec:irxs}

Also known as IR bright galaxies, IR-excess (IRxs) sources are believed to be part of an extreme IR population --- the compton-thick (type-2) AGN --- frequently missed by optical/X-ray surveys. IRxs sources are always selected to have 24\,$\mu$m to $R$-band flux ratio of $f_{24}/f_R>1000$ \citep[except in][who select this type of sources based on a template fit procedure]{Daddi07}. However, highly obscured star-forming galaxies may also present such colors \citep{Dey08,Fiore08,Pope08,Narayanan10}. It has been shown, nevertheless, that the AGN fraction increases as one restricts IRxs samples to brighter fluxes and/or redder NIR-to-optical colors. \citet{Dey08} show that at $f_{24\,\mu{\rm m}}\gtrsim0.6$, $\sim50-80\%$ of the IRxs sample shows power-law SEDs \citep[likely AGN-driven, but see][]{Narayanan10}, and $\sim7-13\%$ have detected X-ray counterparts (a higher fraction than that found for non-IRxs sources). \citet{Donley10} consider 13 IRxs with $f_{24\,\mu{\rm m}}\gtrsim0.7$, which show \emph{Spitzer}-IRS spectra dominated by AGN emission (even though half shows detectable PAH emission), and their X-ray properties hint for heavily obscured (Compton-thick) AGN activity in 80\% of the sample. \citet{Polletta08} considered $f_{24\,\mu{\rm m}}\gtrsim1$ IRxs sources, which also presented extremely red NIR-to-optical colors. All selected sources show the 9.7$\,\mu$m silicate feature in absorption, have extreme MIR luminosities ($L( 6\ \mu \mathrm{m}\,) \simeq 10^{46}\ \mathrm{ergs}\,\ \mathrm{s}\,^{-1}$), and always require the presence of torus emission to explain their MIR SED. Finally, a simpler color combination was explored to fainter fluxes by \citet{Fiore08,Fiore09} and \citet{Treister09b}. Through means of X-ray stacking, the two groups showed that $>80\%$ of IRxs sources with $(R-K)_{\rm Vega}>4.5$ show evidences for heavily obscured (Compton-thick) AGN activity.

The diagnostics considered below rely on optical-to-IR color cuts, more precisely, $R-K$ and $R-[24]$. However, $R$-band photometry is not available in the MUSIC catalog. We thus convert those colors to equivalent ones using $i$-band ($i-K$ and $i-[24]$) considering a power-law spectrum ($f_{\nu}\propto\nu^{\alpha}$). We highlight three criteria. \citet[][D08]{Dey08} select sources with $S_{24}/S_R>1000$ and $S_{24}>300\,\mu{\rm Jy}$ (equivalent to $i-[24]>7$ and $[24]<17.5$). \citet[][F08]{Fiore08} select $S_{24}/S_R>1000$ sources with $R-K>4.5$ ($i-[24]>7$ and $i-K>2.5$\footnote{It should be noted, however, that the $J-K$ color is likely more efficient than $i-K$ in selecting AGN sources \citep[][see also \citealt{Glikman04} for a combination of the two]{Messias10}.}) at $S_{24}>40\,\mu{\rm Jy}$ ($[24]<20$). Finally, we also consider the brightest $S_{24}/S_R>1000$ sources by adopting the flux cut of \citet[][P08]{Polletta08}, $S_{24}>1\,{\rm mJy}$ (corresponding to $[24]<16.5$).

These criteria were applied to the MUSIC and COSMOS catalogs and Table~\ref{tab:irxs} details the numbers of the selected sources by each of the IR color criteria. In spite of the much smaller sample size in CDFS, the results agree remarkably well between the two surveys. In COSMOS, all criteria appear to select most of the brightest objects (P08) in a comparable way (within the error bars), as expected \citep{Donley08,Eckart10}. Regarding the fainter sources, L07 is the most complete, followed by KIM and KI, while S05 and D12 select the fewest IRxs sources.

\begin{deluxetable}{cccc}
\tabletypesize{\scriptsize}
\tablecaption{Selection of IRxs sources.\label{tab:irxs}}
\tablewidth{0pt}
\tablehead{
\colhead{Region} & \colhead{F08} & \colhead{D08} & \colhead{P08}
}
\startdata
\multicolumn{4}{c}{CDFS} \\
\ldots & 77 & 10 & 1 \\
\smallskip
L07 & 72\,(94,15,$_{6}^{3}$) & 9\,(90,41,$_{0}^{10}$) & 1\,(100,100,$_{0}^{0}$) \\ 
\smallskip
S05 & 29\,(38,8,$_{12}^{22}$) & 5\,(50,27,$_{10}^{0}$) & 1\,(100,100,$_{0}^{0}$) \\ 
\smallskip
D12 & 18\,(23,6,$_{9}^{8}$) & 5\,(50,27,$_{0}^{0}$) & 1\,(100,100,$_{0}^{0}$) \\ 
\smallskip
KI  & 40\,(52,10,$_{12}^{12}$) & 7\,(70,34,$_{0}^{0}$) & 1\,(100,100,$_{0}^{0}$) \\ 
\smallskip
KIM & 40\,(52,10,$_{8}^{18}$) & 8\,(80,38,$_{0}^{0}$) & 1\,(100,100,$_{0}^{0}$) \\ 
\hline \\
\multicolumn{4}{c}{{\rm COSMOS}} \\
\ldots & 991 & 256 & 51 \\
\smallskip
L07 & 909\,(92,4,$_{12}^{6}$) & 244\,(95,9,$_{4}^{1}$) & 47\,(92,19,$_{6}^{2}$) \\ 
\smallskip
S05 & 381\,(38,2,$_{11}^{15}$) & 138\,(54,6,$_{4}^{7}$) & 39\,(76,16,$_{6}^{2}$) \\ 
\smallskip
D12 & 202\,(20,2,$_{4}^{7}$) & 128\,(50,5,$_{3}^{4}$) & 41\,(80,17,$_{0}^{2}$) \\ 
\smallskip
KI  & 493\,(50,3,$_{12}^{16}$) & 179\,(70,7,$_{3}^{8}$) & 46\,(90,18,$_{2}^{4}$) \\ 
\smallskip
KIM & 606\,(61,3,$_{18}^{17}$) & 212\,(83,8,$_{9}^{6}$) & 50\,(98,20,$_{0}^{0}$) \\ 
\enddata
\tablecomments{Number of selected IRxs sources. The numbers in parenthesis give the equivalent fractions, the Poisson error, and the photometric upper and lower error bars.}
\end{deluxetable}

\subsection{SDSS QSOs} \label{sec:sdssqso}

QSOs present in the Sloan Digital Sky Survey Quasar Catalogue Data Release 7 \citep[SDSS-DR7,][]{Schneider10} were cross-matched (2'' radius) with the $Spitzer$ IR catalogues from the COSMOS (S-COSMOS), Lockman Hole, ELAIS-N1, and ELAIS-N2 \citep[SWIRE,][]{Lonsdale03} fields using GATOR\footnote{http://irsa.ipac.caltech.edu/applications/Gator/} at IRSA-NASA/IPAC. The final sample is composed by 293 QSOs. $K$-band photometry comes from 2MASS \citep[for 21\% of the sample,][]{Skrutskie06}, UKIDSS-DXS DR8\footnote{UKIDSS uses the UKIRT Wide Field Camera \citep[WFCAM;][]{Casali07} and a photometric system described in \citet{Hewett06}. The pipeline processing and science archive are described in Irwin et al. (in preparation) and \citet{Hambly08}. We have used data from the 8th data release.} \citep[29\%,][]{Lawrence07}, and COSMOS \citep[23\%,][]{Ilbert09}. Overall, there are 186 QSOs with reliable photometry in all IRAC channels. Of these, 140 have also MIPS$_{24\mu\rm{m}}$ photometry, and 142 have $K$-band photometry. We find 107 with full $K$-IRAC-MIPS$_{24\mu\rm{m}}$ coverage. To enhance the high-$z$ regime sampling, we further include 13 SDSS-DR3 QSOs at $z\sim6$ \citep{Jiang06}. Of these, 12 are detected in all IRAC and MIPS$_{24\mu\rm{m}}$ channels, while only five have 2MASS $K$-band data.

\begin{figure}
\plotone{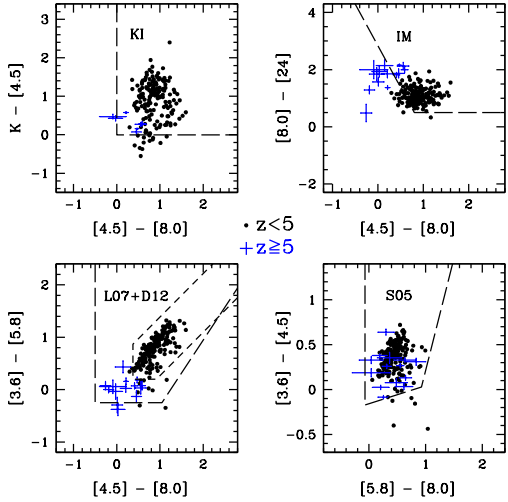}
\caption{The SDSS-DR7 QSOs found in the SWIRE and COSMOS fields together with the \citet{Jiang06} sample displayed in KI (top left), IM (top right), L07 and D12 (bottom left), and S05 (bottom right) color-color spaces. Black dots represent $z<5$ sources (with small average photometric errors), and blue dots (with error bars) otherwise.\label{fig:sdssPlot}}
\end{figure}

Figure~\ref{fig:sdssPlot} shows the location of the QSO sample in the KI, IM (Section~\ref{sec:imhighz}), L07, and S05 color-color spaces. Only sources with reliable photometry in the displayed bands are shown. While L07, S05, and KI select most of the displayed sample ($>90\%$), D12 and KIM select a smaller, yet large, portion of it, respectively, 84\% and 65\%. For $z>5$ QSOs, the IM completeness drops to 36\%, in agreement with Figure~\ref{fig:c24m24}, where QSO templates start to move out of the KIM region at $z\sim4-5$. This is likely a result of both blue rest-frame optical SEDs, which most IRAC bands probe at these redshifts, low metallicity (consequently less dust) in these high redshift sources, and strong H$\alpha$ emission. While blue rest-frame optical SEDs or strong H$\alpha$ emission imply more objects in the blue side of $[4.5]-[8.0]$, low metallicity results in some objects with bluer $[8.0]-[24]$ colors. Note that S05 and D12 retrieve opposite efficiencies in selecting $z>5$ QSOs, where S05 selects all and D12 selects none.

We note, however, that if there is a prior indication for such high redshifts ($z>3$ based on spectroscopy or, e.g., on the LBG technique), then the [8.0]-[24] color can be used by itself and much more efficiently for the identification of AGN (cf. Figure~\ref{fig:z824}). For $z>5$ QSOs, all but one show $[8.0]-[24]>1$. The few QSOs with blue $K-[4.5]$ (top left panel) colors are explained in light of the discussion in Section~\ref{sec:ki}: these are potentially AGN with smaller optical-to-IR flux ratios and/or sources possessing strong line emission.

The high completeness levels achieved with this optical selected sample show the eclectic selection of IR criteria. However, optically selected AGN are not the main targets of IR AGN diagnostics, as, by definition, optical surveys \textit{do} detect them. The most interesting use of these criteria is to recover sources undetected at X-ray and optical wavelengths. Sections~\ref{sec:irxs} (above) and \ref{sec:hzrg} (below) are, in this respect, much more representative of the usefulness of IR AGN diagnostics.

\subsection{High redshift Radio Galaxies} \label{sec:hzrg}

To test yet another AGN population, we now consider High-\textit{z} Radio Galaxies (H\textit{z}RGs). These are among the most luminous sources in the Universe and are believed to host powerful AGN. We use the sample of 71 H\textit{z}RGs from \citet{Seymour07}. These are all at $z>1$, a redshift range where no normal galaxy is believed to contaminate the AGN IM region proposed in Section~\ref{sec:imhighz}. This is a classic example --- such as that of LBGs or spectroscopically confirmed galaxies at $z>1$ --- for the direct application of the IM boundaries. Having this, the $K-[4.5]>0$ color cut is not required to disentangle AGN/non-AGN dominated sources at $z<1$ (see Section~\ref{sec:ki}), meaning that one may consider $[4.5]-[8.0]$ and $[8.0]-[24]$ colors alone to determine whether AGN or stellar emission dominates the IR spectral regime.

\begin{figure}
\plotone{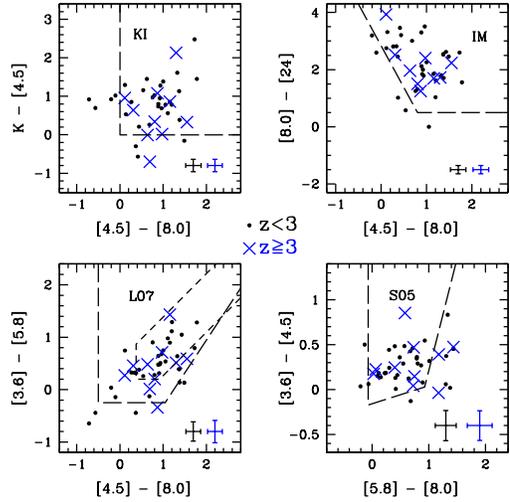}
\caption{The HzRG ($z>1$) sample from \citet{Seymour07} displayed in the same color-color spaces as in Figure~\ref{fig:sdssPlot}. Note the objects at $2<[8.0]-[24]<4$ which are even redder than QSOs (Figure~\ref{fig:sdssPlot}). Dots show the $z<3$ population, crosses show the population at $z\geq3$. The average photometric errors of this sample are shown on the bottom right of each panel for the $z<3$ (left error bars) and $z\geq3$ (right) samples.\label{fig:hzrgIM}}
\end{figure}

Figure~\ref{fig:hzrgIM} shows the location of 62 H\textit{z}RGs in the IM color-color diagram (top right panel). Note the difference to SDSS QSOs (Figure~\ref{fig:sdssPlot}), where H\textit{z}RGs show predominantly redder $[8.0]-[24]$ colors. The AGN region correctly selects as AGN 88\% (37 sources) of the sample with adequate photometry (42 sources detected at 4.5, 8.0, and 24$\mu$m). If no redshift estimate was available, however, one would need the $K-[4.5]>0$ color cut to apply the IM AGN criterion, i.e., the KIM criterion. The application of KIM would result in a 78\% completeness level (28 out of 36 sources). The remaining panels show that L07 selects 86\% (38 out of 44 sources), D12 selects 27\% (12 out of 44 sources), S05 selects 73\% (32 out of 44 sources), and KI selects 78\% (32 out of 41 sources).

Again we emphasize that much of the improvement of KI/KIM over the commonly used L07 and S05 diagnostics is in terms of reliability, an indicator which is not being evaluated with this sample nor those in Sections~\ref{sec:irxs} and \ref{sec:sdssqso}.

\subsection{Sub-Millimetre Galaxies}

In \citet{Hainline11} one can find estimates for the non-stellar contribution (shown in that work to be related to AGN activity) to the IR SED of Sub-Millimetre Galaxies (SMGs), showing that significant AGN emission is present even at $1.6\,\mu$m ($\sim50-60\%$ of the sample with an estimate shows AGN contributions of $\gtrsim20-10\%$ at these wavelengths). The estimates for the non-stellar contribution were obtained photometrically by fitting the SMGs SEDs with a combination of pure star-forming models and power-law spectra ($f\propto\nu^\alpha$, with $\alpha=-2$ or $-3$). These non-stellar contribution estimates depend on whether one considers a constant star-formation history or an initial instant burst followed by passive evolution. These assumptions provide two extreme scenarios, and here we adopt the average between the two. The sample is that from \citet{Chapman05}, found to be at $0.1\lesssim{z_{\rm spec}\lesssim3.4}$ ($\sim85\%$ of the sample is at $z_{\rm spec}>1$). The $K$-band photometry is presented in \citet{Smail04}, while the IRAC and MIPS photometry is presented in \citet{Hainline09}.

\begin{figure}
\plotone{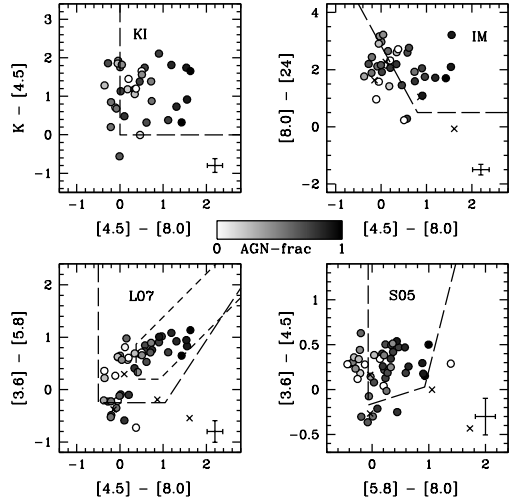}
\caption{The SMG sample from \citet{Chapman05} displayed in the same color-color spaces as in Figure~\ref{fig:sdssPlot}. Only sources detected in the referred bands in each panel are displayed. The average photometric errors are shown on the bottom right of each panel. The data point intensity shows the AGN fraction at observed $8\,\mu$m (intensity bar in the center). Sources with AGN contribution upper limits are shown as crosses.\label{fig:smg}}
\end{figure}

Figure~\ref{fig:smg} shows the distribution of SMGs in different color-color spaces. The intensity of each data point relates to the AGN contribution at observed $8\,\mu$m (the darker it is, the higher is the AGN contribution). Specifically in the IM plot (upper right panel), even though there are two sources with an estimated 0\% contribution from AGN inside the selection region, all but one (with 25\%) of the remainder (19 sources) have $\geq50\%$ of AGN contribution. Overall, the IM selected sample shows an average of $62^{+3}_{-5}\%$\footnote{The errors only consider the propagated errors associated to each individual estimate used for averaging.} of AGN contribution at observed $8\,\mu$m. L07, S05, D12, and KI show\footnote{Separately considering sources detected in bands involved in each criteria, hence subject to selection effects. For instance, 49 SMGs are detected in all four IRAC channels (used for L07, S05, and D12 statistics), while 34 are detected in $K$-band, 4.5\,$\mu$m and 8.0\,$\mu$m (used for KI statistics).}, respectively, averages of $55^{+2}_{-3}\%$, $66^{+2}_{-4}\%$, $75^{+2}_{-7}\%$, and $60^{+3}_{-5}\%$. Although all criteria select at least one SMG with an estimated 0\% contribution from AGN, the average AGN contribution is always $\geq55\%$. This means that for many SMGs, although the AGN does not dominate the IR emission, the AGN contributes significantly to the IR SED of the galaxy. The same is expected for IR selected AGN, this is, although we see selected sources where the AGN does not dominate at IR wavelengths, the AGN is present and active, contributing enough for the source to be selected as an IR AGN.

\section{Implications for JWST surveys} \label{sec:impjwst}

The start of scientific observations of \emph{JWST}, the successor of $Spitzer$ at mid-IR wavelengths, is expected for 2018. It will be a 6.5\,m space telescope which will operate from 0.6 to 27\,${\mu}$m. As highlighted in this and previous work, this spectral regime has great potential for separating AGN from normal (non-AGN) galaxies.

The sensitivity will of course be better than ever before, and the high-\textit{z} universe will be probed with unprecedented detail. Many galaxies will be studied with MIR spectroscopy, and signs of AGN activity will be naturally found that way \citep[e.g.,][]{Laurent00,Armus07,Veilleux09,Fu10}. When dealing with large surveys, however, with thousands of sources and many close to the detection limit, AGN selection will have to rely on photometric diagnostics such as the KI/KIM criteria presented here. By selecting AGN candidates over a broad range of redshifts, $0<z<7$, the KIM criterion will enable the study of AGN phenomena to the earliest epochs.

While the KI/KIM criteria can already be applied to current data from $Spitzer$, potentially more efficient MIR criteria will be possible with the large wavelength coverage of the \emph{JWST}. Using planned \emph{JWST} filter response curves\footnote{Provided online at:\\ http://www.stsci.edu/jwst/instruments/\\nircam/instrumentdesign/filters/index\_html \\http://www.stsci.edu/jwst/instruments/miri/\\instrumentdesign/miri\_glance.html .}, we suggest a possible and promising color-color space alternative to that proposed in Section~\ref{sec:imhighz}, using the MIRI 10$\mu$m and 21${\mu}$m filters instead of the IRAC 8.0$\mu$m and MIPS 24$\mu$m bands, and the NirCAM 4.4$\mu$m instead of IRAC 4.5$\mu$m \citep[note that these are bands close to those used in Wide-field IR Survey Explorer, WISE; see also][]{Assef10,Stern12}. The selection conditions are the following: \[\left[10\right]-\left[21\right] > -5.2\times(\left[4.4\right]-\left[10\right])+2.2\] and \[\left[10\right]-\left[21\right] > 0.1\]

\begin{figure}
\plotone{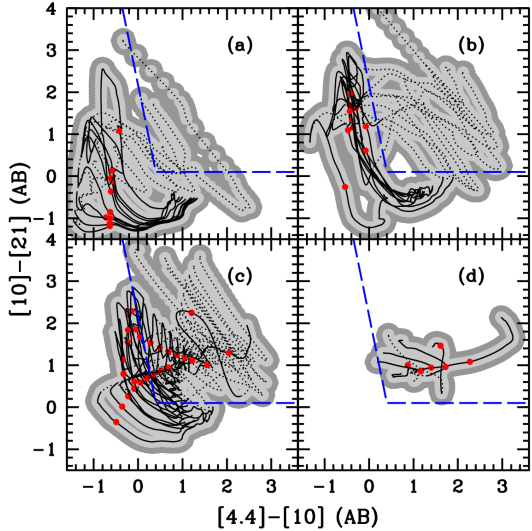}
\caption{An alternative color-color space with \textit{JWST} bands which might improve the AGN selection at $0<z<7$. Symbols and panels definition as in Figure~\ref{fig:c1324}. The light-/dark-grey error circles in the tracks result from larger colour gradients between redshift steps ($\Delta z=0.01$ each). These ``circles" correspond to the error ellipse of each redshift point in the colour-tracks.\label{fig:jwst}}
\end{figure}

\begin{figure}
\plotone{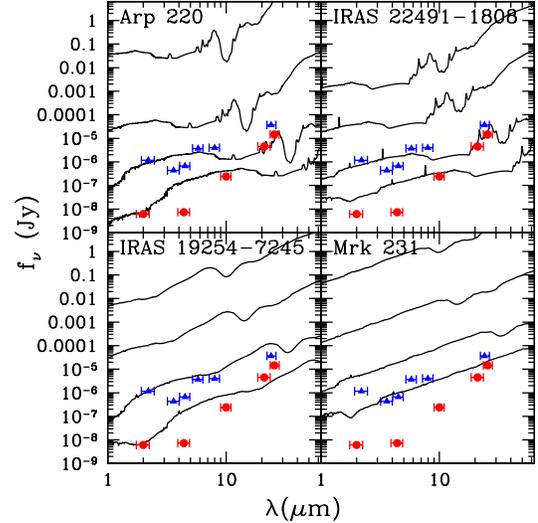}
\caption{SED flux evolution with redshift for two star-formation dominated systems (Arp220 and IRAS 22491-1908, upper panels), and two AGN dominated systems (IRAS 19254-7245 and Mrk231, lower panels). The redshift steps are $z=z_0,\,0.5,\,2.5,\,7$. The blue dots indicate the $K$-IRAC-MIPS$_{24\mu{\rm m}}$ CDFS $10\sigma$ total flux level \citep[based in Table 1 of][]{Wuyts08}, red dots give the $10\sigma$ level (at equivalent CDFS integration times) of the \textit{JWST} filters: 2.0$\,\mu$m, 4.4$\,\mu$m, 10$\,\mu$m, 21$\,\mu$m, and 25$\,\mu$m. At longer wavelengths, the gap between \textit{Spitzer} and \textit{JWST}'s sensitivities is smaller due to the warmer telescope thermal background of \textit{JWST}.\label{fig:jwstTp}}
\end{figure}

\begin{figure}
\plotone{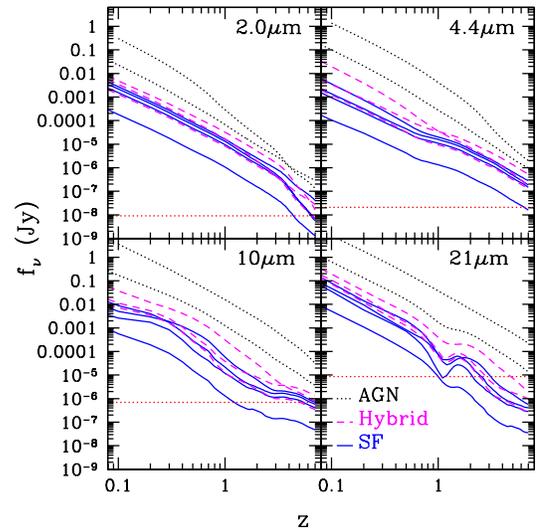}
\caption{Flux evolution with redshift for starbursts (blue solid line), hybrids (magenta dashed line), and AGN (black dotted line) in four \textit{JWST} filters. Red dotted horizontal lines mark the $10\sigma$ level (10,000$\:$s integration) of each filter.\label{fig:jwstFt}}
\end{figure}

In Figure~\ref{fig:jwst}, the four panels show that the [4.4]-[10] versus [10]-[21] color-color space seems to present a better selection of the AGN/Hybrid model tracks. The AGN model tracks are better delineated by the selection boundaries (recovering the $z\gtrsim4-5$ QSO missed by KIM) and, as a bonus, the 21$\mu$m filter is over three times more sensitive than the planed MIRI 25$\mu$m filter (equivalent to the MIPS 24$\mu$m filter), increasing the probability of a detection needed for an AGN classification. This is shown in Figures~\ref{fig:jwstTp} and \ref{fig:jwstFt}, where AGN dominated sources are detected up to the highest redshift considered in this work ($z\sim7$).

Finally, the improved resolution of JWST will allow the study of the nuclear region alone up to higher redshifts. This, of course, allows the identification of lower luminosity AGN \citep[e.g.,][]{Honig10,Asmus11}, thus overcoming the difficulty in selecting such systems in the IR.

\section{Conclusions} \label{sec:conc}

Based on semi-empirical galaxy SED templates, we have developed physically motivated IR color criteria for the selection of a wide variety of AGN in a large redshift range ($0<z<7$). As well as the application to existing data \citep[e.g., the recently available WISE data\footnote{http://wise2.ipac.caltech.edu/docs/release/prelim/} at 3.4, 4.6, 12, and 22\,$\mu$m,][]{Wright10}, these criteria are particularly relevant for the \emph{JWST}, given the wide MIR spectral range considered. We thus propose new AGN IR diagnostics. The $K-[4.5]$ color is ideal for the $z<1$ universe (Section~\ref{sec:ki}); KI is a reliable alternative to the IRAC-based diagnostics at $z\lesssim2.5$ (based on Poisson errors alone, Sections~\ref{sec:ki} and \ref{sec:goodssample}); and KIM (Section~\ref{sec:imhighz}), a four band (K, 4.5, 8.0, and 24\,${\mu}$m), three color (K-[4.5], [4.5]-[8.0], and [8.0]-[24]) criterion is proposed as a new `wedge' criterion to select AGN hosts over the full $0<z<7$ range, based on template colour tracks alone (the current control samples prevent any conclusion at $z\gtrsim2.5$ due to small sample statistics).

The \textit{absolute} efficiency of IR AGN criteria is, at this stage, impossible to obtain as the control (X-ray and optical selected) samples used here are themselves incomplete, hence, results should be treated with care. For instance, KI achieves reliabilities of $\sim50\%$ in CDFS ($\sim$90\% in COSMOS) and KIM shows reliabilities of $\sim$60--80\% ($\sim$90\%). These should be considered lower-limits, because X-ray and spectroscopic data may not have been deep enough to reveal AGN features (Section~\ref{sec:ctrlsp}), and we show that the selected sources with SF-dominated SEDs, are still likely to host an active AGN.

Nevertheless, in the coming years, these criteria should be improved as a result of the rich variety of filters to be incorporated in the instruments on board the upcoming \textit{JWST} (Section~\ref{sec:impjwst}). The ability to track AGN activity since the end of the reionization epoch will hold great advantages for the study of galaxy-AGN co-evolution.

\acknowledgments
The authors thank the two anonymous referees for substantial and constructive comments that have led to relevant and significant improvements in this paper, and would also like to thank useful comments and suggestions by Jessica Krick and Pablo P\'erez-Gonz\'alez. HM acknowledges the frequent use of Topcat and VOdesk. HM acknowledges the support from Funda\c{c}\~{a}o para a Ci\^{e}ncia e a Tecnologia through the scholarship SFRH/BD/31338/2006. HM and JA acknowledge support from Funda\c{c}\~{a}o para a Ci\^{e}ncia e a Tecnologia through the projects PTDC/CTE-AST/105287/2008 and PEst-OE/FIS/UI2751/2011. HM acknowledges the support by UCR while visiting Dr. Bahram Mobasher as a visitor scholar. MS acknowledges support by the German Deutsche Forschungsgemeinschaft, DFG Leibniz Prize (FKZ HA 1850/28-1).

\begin{appendix}

\begin{figure}
\plottwo{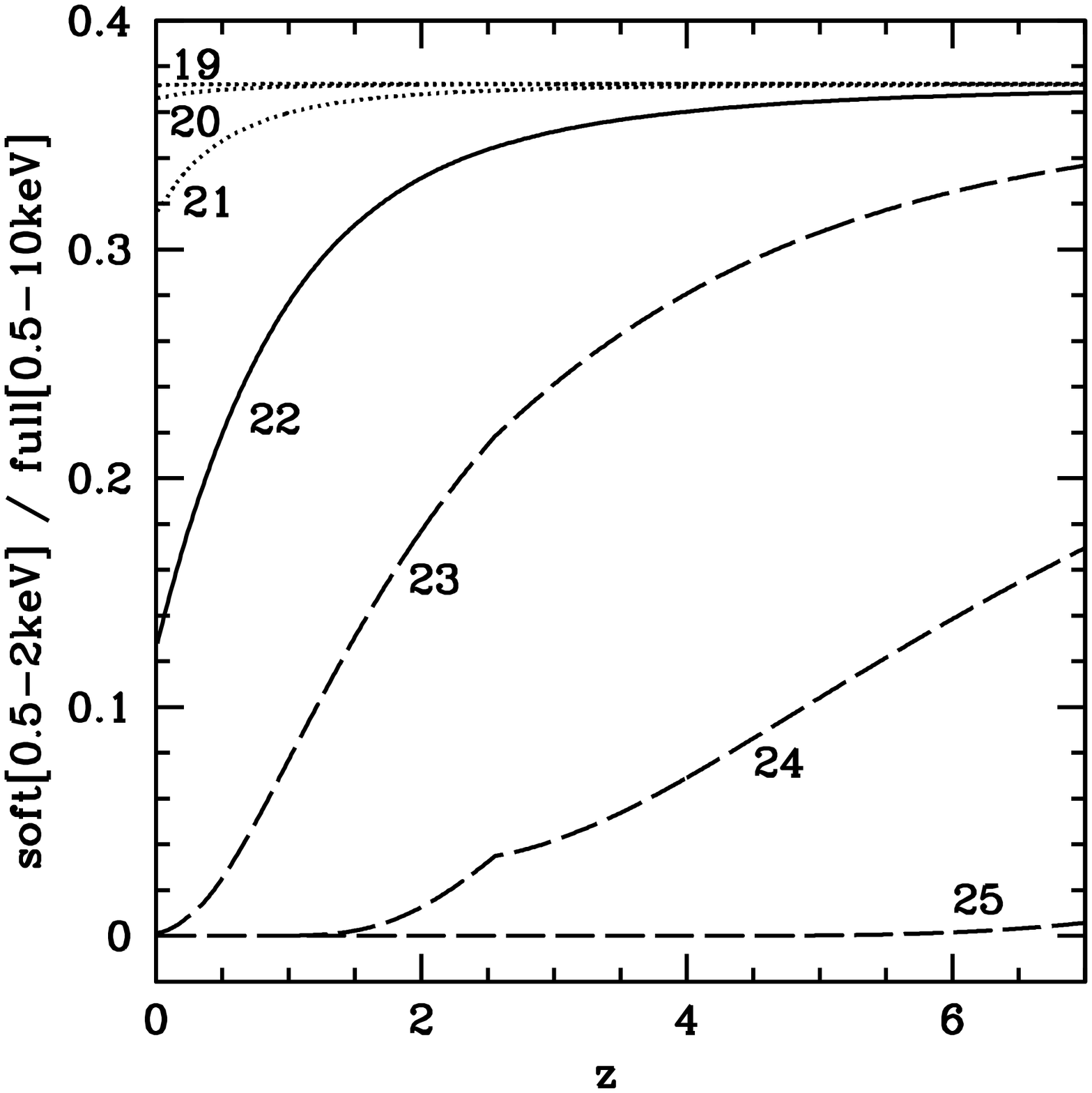}{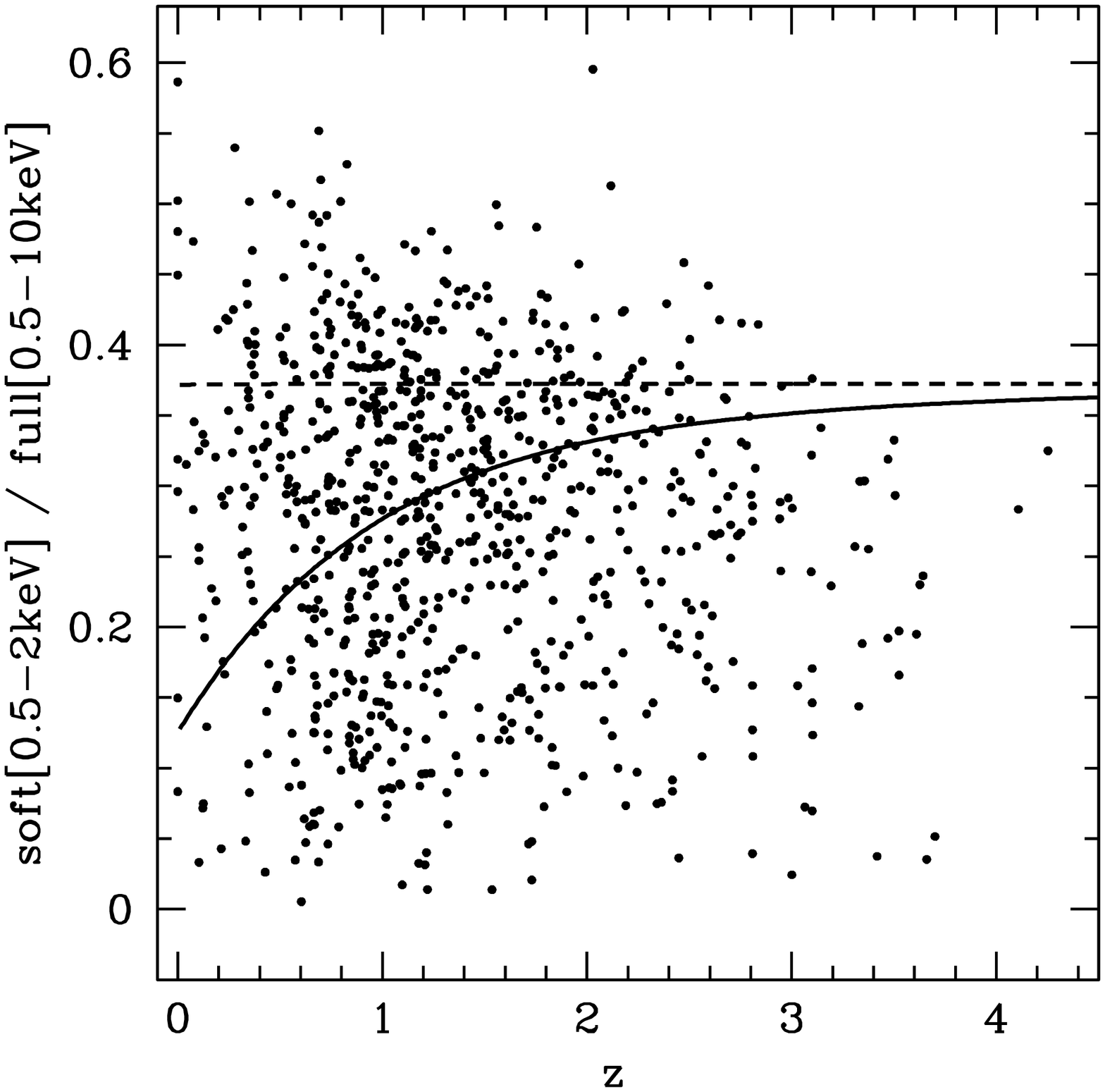}
\caption{The left hand-side panel shows the dependency of the flux ratio on column density ($\rm{N_H}$) and redshift. The numbers refer to the $\rm{\log(N_H[cm^{-2}])}$ value corresponding to each line. The reference level $\rm{\log(N_H[cm^{-2}])}=22$ is shown as solid line. The right-hand panel shows the observed flux ratios for the CDFS and COSMOS samples. The $\rm{\log(N_H[cm^{-2}])}=22$ value is shown again as solid line for reference, as well as the $\rm{\log(N_H[cm^{-2}])}=19$ value, above which we have considered $\rm{\log(N_H[cm^{-2}])}=0$ and the observed photon index to estimate X-ray luminosities.\label{fig:fratz}}
\end{figure}

It has been pointed out that the estimate of column densities ($\rm{N_H}$) based on soft-to-hard X-ray flux ratios overestimates the type-2 (obscured) object fraction with increasing redshift \citep{Ueda03,Akylas06,Donley12}. The fact that we consider soft-to-full or hard-to-full band ratios does not change the scenario. This bias effect happens due to a harder spectral range being probed at higher redshifts by the observed 0.5--8/10\,keV spectral window plus the error associated with the flux measurement. Figure~\ref{fig:fratz}a shows the soft-to-full ratio variation depending on redshift and $\rm{N_H}$. One can see that with increasing redshifts, for the smallest $\rm{N_H}$ values, the associated flux ratios become problematically degenerate\footnote{At $z\sim0$, there is also a degeneracy at $\rm{\log(N_H[cm^{-2}])}=23-25$, but it does not represent a problem, because the constraint considered to distinguish type-1 from type-2 objects is $\rm{\log(N_H[cm^{-2}])}=22$.}. Would there be perfect precision and accuracy in the flux measurements, this would hold no problem, however, there is a flux error which scatters the real values producing the observational bias. Figure~\ref{fig:fratz}b shows the observed values from the CDFS and COSMOS samples for reference.

\begin{figure}
\plottwo{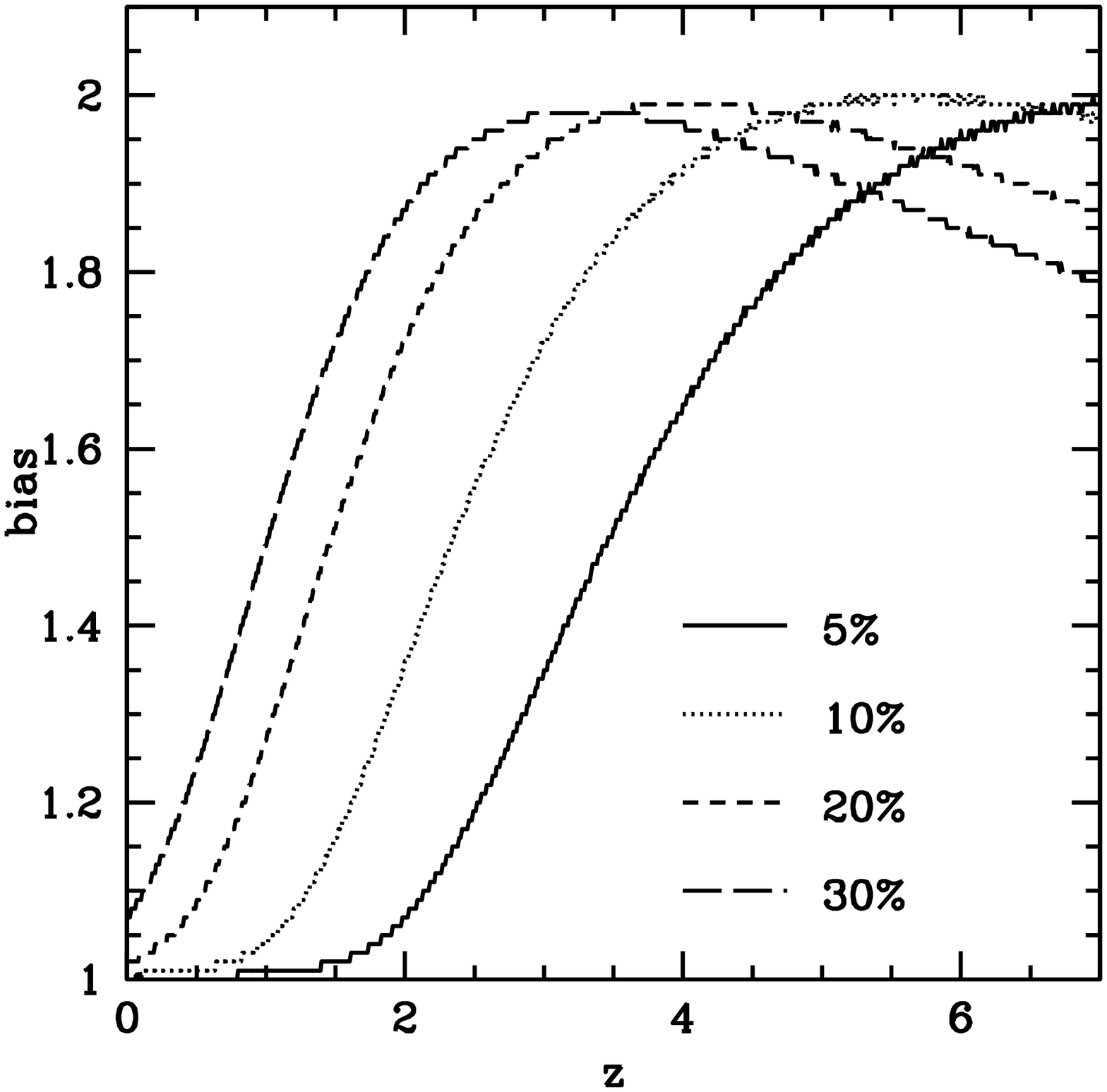}{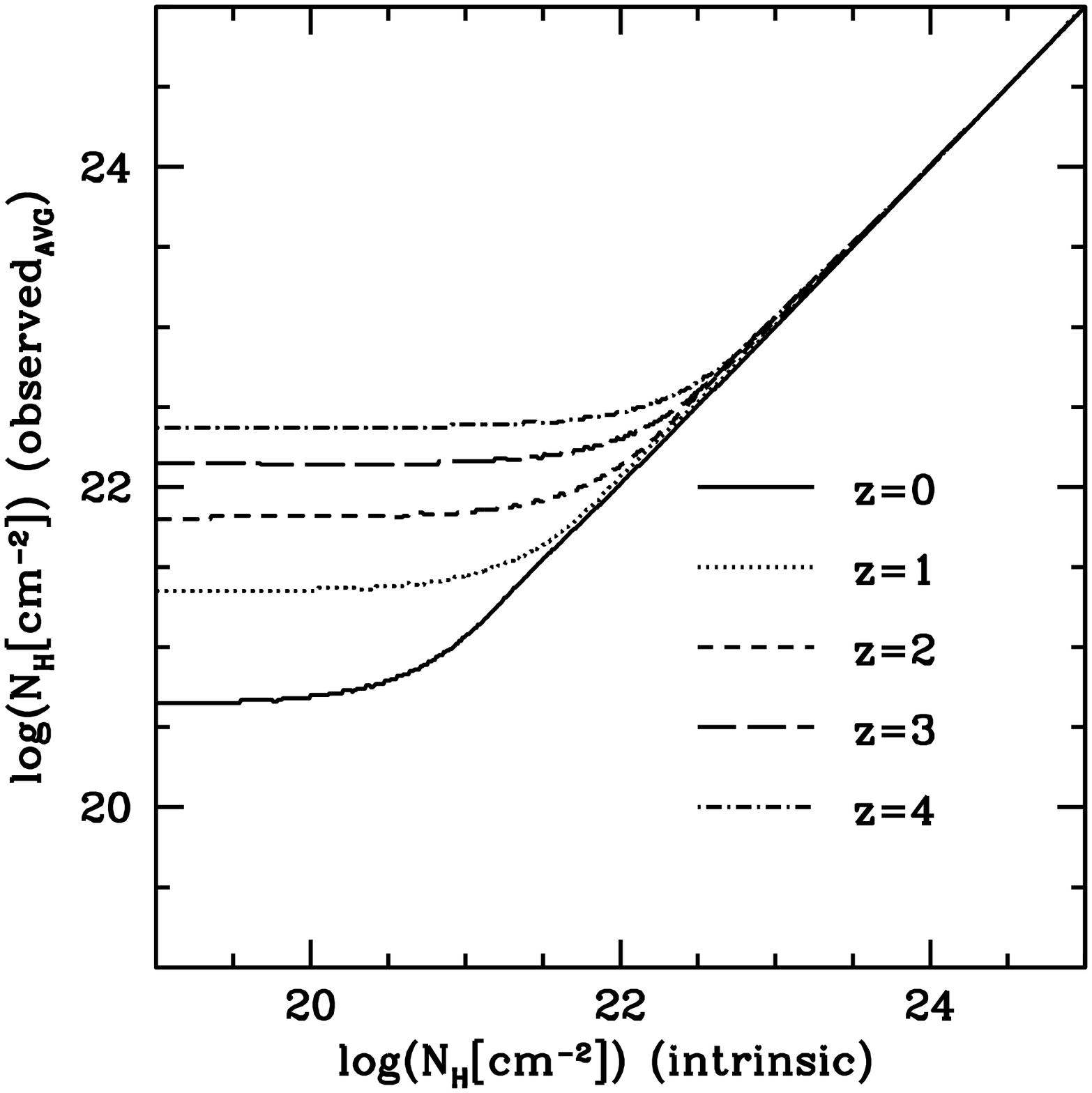}
\caption{The left hand-side panel (a) shows the bias toward type-2 objects evolution with redshift for different levels of error in the flux ratio estimate. The right hand-side panel (b) shows, for a 20\% error level in the flux ratio estimate, the average overestimate depending on column density ($\rm{N_H}$) and redshift.\label{fig:nhbias}}
\end{figure}

Figure~\ref{fig:nhbias}a shows the overall bias toward type-2 objects over type-1 objects depending on redshift and flux ratio error (in percent). In order to produce this figure, we have considered an initial 50/50\% fraction of type-1/type-2 objects and a Gaussian probability density distribution (PDD) centred at flux ratios corresponding to $19<\rm{\log(N_H[cm^{-2}])}<25$ and with a $\sigma=5,\,10,\,20,\,30\%$ of the flux ratio. The Gaussian PDD gives the probability for the output (observed) classification to be \emph{unobscured} ($\rm{\log(N_H[cm^{-2}])}\leq22$) or \emph{obscured} ($\rm{\log(N_H[cm^{-2}])}>22$), consequently providing the overall output bias shown in Figure~\ref{fig:nhbias}a. In this figure, the bias is calculated as the ratio between the output (observed) type-2/type-1 ratio and the input (intrinsic) type-2/type-1 ratio. Note the trends maximize at $\sim2$. If we instead had restricted this exercise to the $20<\rm{\log(N_H[cm^{-2}])}<24$ range \citep[considered in][]{Donley12}, the maximum value would be at $\sim1.8$. This is the bias found by \citet{Donley12} at $z\sim3$. This means, for the $z\sim3$ XMM sources used in that work, the flux ratios are affected on average by a $\sim30\%$ error.

The Gaussian PDD also provides the average overestimate associated with a given intrinsic $\rm{N_H}$ value depending on flux ratio error and redshift. This is shown in Figure~\ref{fig:nhbias}b for a 20\% flux ratio error. These trends allow us to attempt the correction of the referred bias, where the observed $\rm{N_H}$ value (y-axis in Figure~\ref{fig:nhbias}b) corresponds to a intrinsically smaller $\rm{N_H}$ value (x-axis) for a given flux ratio error at the source's redshift. The flux ratios used to estimate $\rm{N_H}$ in this work are all based on measurements with errors less than a third of the fluxes, a propagated flux ratio error of $<47\%$ (the average accepted flux ratio error was $\sim19\%$). Even though individually the corrected value \emph{should not} be regarded as the best estimate, this average correction applied to the overall population will diminish the bias. The number of type-1 objects which would be classified as type-2 if no correction was applied is small: 14 in CDFS and 9 in COSMOS. Most of these (19 objects) are at $z\gtrsim3$.

\end{appendix}

\vspace{0.5cm}

\twocolumngrid

\end{document}